\documentclass[prl,aps,epsfig,superscriptaddress,twocolumn]{revtex4-1}
\usepackage{amsfonts,amssymb,amscd,amsthm}
\usepackage{graphicx}
\usepackage{mathrsfs}
\usepackage[intlimits]{amsmath}
\usepackage[colorlinks, citecolor=red]{hyperref}
\usepackage{etoolbox}
\usepackage{upgreek}
\usepackage{braket}

\begin{document}

\title{Individually Addressed Entangling Gates in a Two-Dimensional Ion Crystal}

\author{Y.-H. Hou}
\thanks{These authors contribute equally to this work}%
\affiliation{Center for Quantum Information, Institute for Interdisciplinary Information Sciences, Tsinghua University, Beijing 100084, PR China}

\author{Y.-J. Yi}
\thanks{These authors contribute equally to this work}%
\affiliation{Center for Quantum Information, Institute for Interdisciplinary Information Sciences, Tsinghua University, Beijing 100084, PR China}

\author{Y.-K. Wu}
\thanks{These authors contribute equally to this work}%
\affiliation{Center for Quantum Information, Institute for Interdisciplinary Information Sciences, Tsinghua University, Beijing 100084, PR China}
\affiliation{Hefei National Laboratory, Hefei 230088, PR China}

\author{Y.-Y. Chen}
\affiliation{Center for Quantum Information, Institute for Interdisciplinary Information Sciences, Tsinghua University, Beijing 100084, PR China}

\author{L. Zhang}
\affiliation{Center for Quantum Information, Institute for Interdisciplinary Information Sciences, Tsinghua University, Beijing 100084, PR China}

\author{Y. Wang}
\affiliation{Center for Quantum Information, Institute for Interdisciplinary Information Sciences, Tsinghua University, Beijing 100084, PR China}
\affiliation{HYQ Co., Ltd., Beijing 100176, PR China}

\author{Y.-L. Xu}
\affiliation{Center for Quantum Information, Institute for Interdisciplinary Information Sciences, Tsinghua University, Beijing 100084, PR China}

\author{C. Zhang}
\affiliation{Center for Quantum Information, Institute for Interdisciplinary Information Sciences, Tsinghua University, Beijing 100084, PR China}
\affiliation{HYQ Co., Ltd., Beijing 100176, PR China}

\author{Q.-X. Mei}
\affiliation{HYQ Co., Ltd., Beijing 100176, PR China}

\author{H.-X. Yang}
\affiliation{HYQ Co., Ltd., Beijing 100176, PR China}

\author{J.-Y. Ma}
\affiliation{HYQ Co., Ltd., Beijing 100176, PR China}

\author{S.-A. Guo}
\affiliation{Center for Quantum Information, Institute for Interdisciplinary Information Sciences, Tsinghua University, Beijing 100084, PR China}

\author{J. Ye}
\affiliation{Center for Quantum Information, Institute for Interdisciplinary Information Sciences, Tsinghua University, Beijing 100084, PR China}

\author{B.-X. Qi}
\affiliation{Center for Quantum Information, Institute for Interdisciplinary Information Sciences, Tsinghua University, Beijing 100084, PR China}

\author{Z.-C. Zhou}
\affiliation{Center for Quantum Information, Institute for Interdisciplinary Information Sciences, Tsinghua University, Beijing 100084, PR China}
\affiliation{Hefei National Laboratory, Hefei 230088, PR China}

\author{P.-Y. Hou}
\affiliation{Center for Quantum Information, Institute for Interdisciplinary Information Sciences, Tsinghua University, Beijing 100084, PR China}
\affiliation{Hefei National Laboratory, Hefei 230088, PR China}

\author{L.-M. Duan}
\email{lmduan@tsinghua.edu.cn}
\affiliation{Center for Quantum Information, Institute for Interdisciplinary Information Sciences, Tsinghua University, Beijing 100084, PR China}
\affiliation{Hefei National Laboratory, Hefei 230088, PR China}
\affiliation{New Cornerstone Science Laboratory, Beijing 100084, PR China}

\begin{abstract}
Two-dimensional (2D) ion crystals have become a promising way to scale up qubit numbers for ion trap quantum information processing. However, to realize universal quantum computing in this system, individually addressed high-fidelity two-qubit entangling gates still remain challenging due to the inevitable micromotion of ions in a 2D crystal as well as the technical difficulty in 2D addressing. Here we demonstrate two-qubit entangling gates between any ion pairs in a 2D crystal of four ions. We use symmetrically placed crossed acousto-optic deflectors (AODs) to drive Raman transitions and achieve an addressing crosstalk error below $0.1\%$. We design and demonstrate a gate sequence by alternatingly addressing two target ions, making it compatible with any single-ion addressing techniques without crosstalk from multiple addressing beams. We further examine the gate performance versus the micromotion amplitude of the ions and show that its effect can be compensated by a recalibration of the laser intensity without degrading the gate fidelity. Our work paves the way for ion trap quantum computing with hundreds to thousands of qubits on a 2D ion crystal.
\end{abstract}

\maketitle

Quantum computation has attracted wide research interest because of its potential computational power beyond the framework of classical computers \cite{nielsen2000quantum}. To date, various physical platforms have demonstrated elementary quantum operations whose fidelities are above the threshold of fault-tolerant quantum computing \cite{Acharya2023,PhysRevLett.129.010502,PhysRevLett.117.060504,PhysRevLett.117.060505,Bluvstein2024,Rong2015,Noiri2022}. However, to obtain a large-scale error-corrected universal quantum computer and to solve practical problems like factoring, the currently available qubit number of tens to hundreds \cite{Acharya2023,PhysRevLett.127.180501,PhysRevX.13.041052,chen2023benchmarking,Bluvstein2024} still needs to be improved by several orders of magnitude \cite{PhysRevA.86.032324}, which remains an outstanding challenge to the community.

As one of the leading physical platforms for quantum information processing, trapped ions are remarkable for their high-fidelity quantum operations and long-range entangling gates \cite{10.1063/1.5088164}. A single-qubit gate fidelity above $99.9999\%$ \cite{PhysRevLett.113.220501}, a two-qubit gate fidelity above $99.9\%$ \cite{PhysRevLett.117.060504,PhysRevLett.117.060505,PhysRevLett.127.130505}, and a state-preparation-and-measurement fidelity above $99.99\%$ \cite{PhysRevLett.129.130501} have been reported. However, the number of qubits in the commonly used one-dimensional (1D) configuration of ion crystals is seriously limited \cite{wineland1998experimental,PhysRevLett.77.3240,clark2001proceedings}. To further scale up the qubit number, one promising scheme is the quantum charge-coupled device (QCCD) architecture where ions are physically shuttled into small crystals in separated regions for different tasks like storage, quantum gates and measurements \cite{wineland1998experimental,Kielpinski2002,PhysRevX.13.041052}. However, currently this scheme is limited by the relatively slow transport speed of the ions and the follow-up cooling operations, which take up more than $98\%$ of the time budget \cite{PhysRevX.13.041052}. On the other hand, an ion-photon quantum network can connect up individual quantum computing modules \cite{10.5555/2011617.2011618,RevModPhys.82.1209,PhysRevA.89.022317} and is compatible with the 1D ion crystals or the QCCD architectures. However, its performance is restricted by the relatively low efficiency for ion-photon entanglement generation and connection \cite{PhysRevLett.124.110501}. Therefore it is still desirable to push up the qubit number in each module to reduce the communication overhead.

Recently, 2D ion crystals have been used to largely extend the ionic qubit number in a single Paul trap \cite{Szymanski2012crystal,https://doi.org/10.1002/qute.202000068,Xie_2021,PhysRevA.105.023101,PRXQuantum.4.020317,Qiao2024,guo2023siteresolved}. In particular, quantum simulation with up to $300$ ions has been demonstrated \cite{guo2023siteresolved}.
Despite the site-resolved single-shot measurement capability, previous experiments are still restricted to global quantum manipulations, while individually addressed single-qubit and two-qubit quantum gates have not yet been realized. Fundamentally, the inevitable micromotion of ions in a 2D crystal seems to affect the gate fidelity, although theoretical works already show that the micromotion is a coherent process and can in principle be included into the gate design \cite{PhysRevA.90.022332,wang2015quantum,Bermudez_2017,PhysRevA.101.052332,PhysRevA.103.022419}. Technically, individual addressing in 2D is also more complicated than 1D \cite{patent-ZL202110047218.2,pu2017experimental,PhysRevLett.125.150505,Barredo2018,Ebadi2021}. For example, although crossed AODs can address a rectangular array by generating multiple beams within a single row or a single column \cite{patent-ZL202110047218.2}, it struggles to address two ions along the diagonal without creating undesired light spots on the other two corners. Here we solve this problem by developing a two-qubit gate sequence that addresses only one ion at a time. We demonstrate a crosstalk below $0.1\%$ by symmetrically placed crossed AODs, and realize two-qubit entangling gates between any ion pairs in a 2D crystal. By pushing an ion pair away from the RF null axis of the trap, we adjust the micromotion amplitude and experimentally show that it does not affect the gate fidelity up to a recalibration of the laser intensity. Combined with gate sequences with more degrees of freedom, our work can be readily applied to larger 2D ion crystals, thus paves the way toward ion trap quantum computing with hundreds to thousands of qubits in a single quantum computation module.

\begin{figure}[!tbp]
	\centering
	\includegraphics[width=\linewidth]{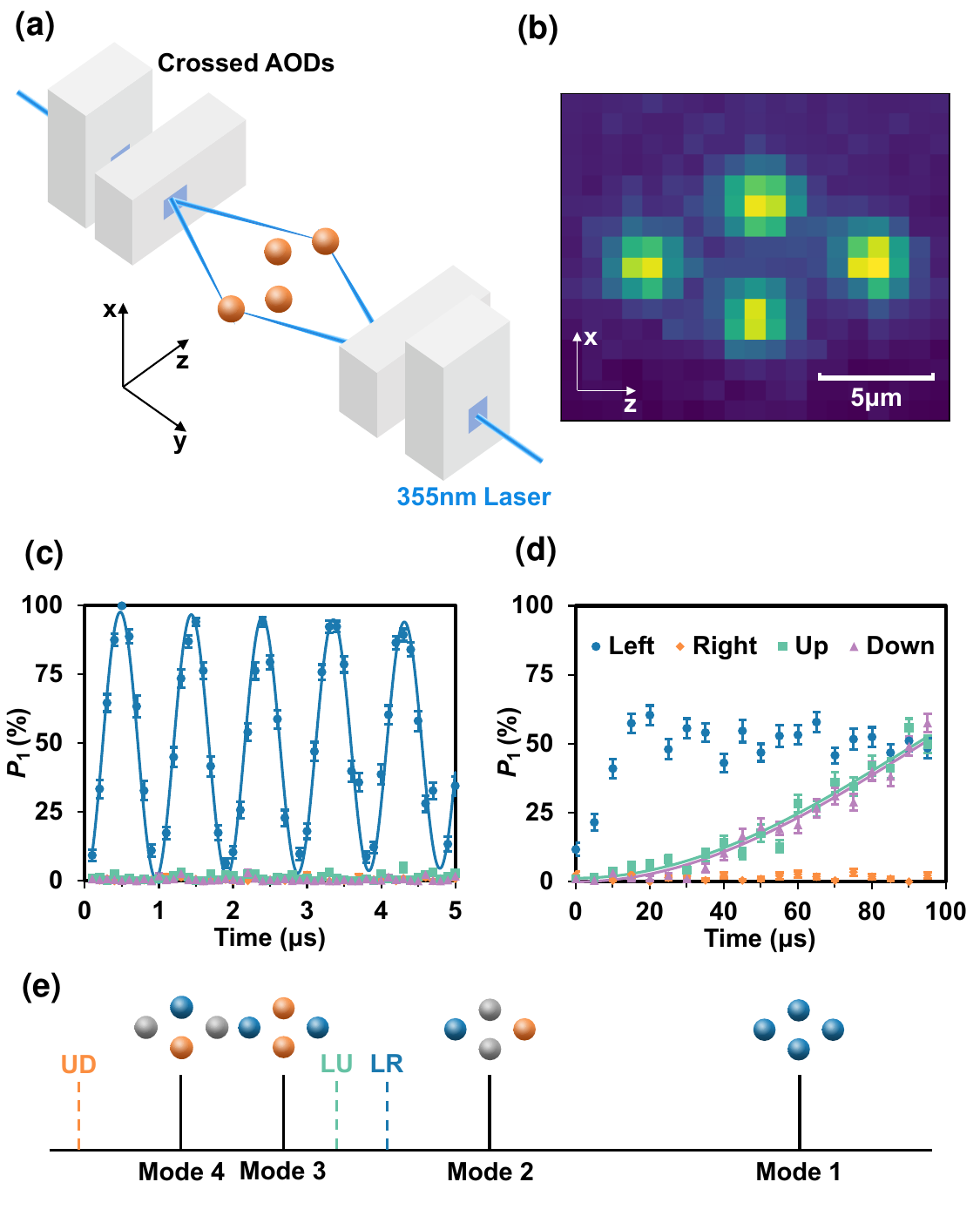}
	\caption {Experimental setup and 2D individual addressing. (a) We address a 2D ion crystal in the $xz$ plane by two pairs of symmetrically placed crossed AODs. By tuning the driving frequency of the AODs, the incoming $355\,$nm laser can be steered in the $x$ or $z$ directions to form a Raman transition on a target ion. By applying two frequencies to an AOD, we can generate addressing beams for two target ions simultaneously, e.g., along the $z$ direction.
(b) The image of the four-ion crystal on an EMCCD camera. The $z$ axis is chosen to be the micromotion-free axial direction of the trap. Below we label the four ions by left, right, up and down according to this image.
(c) Rabi oscillation of the four ions when addressing the left ion. (d) Similar plot as (c) for longer evolution time. Crosstalk infidelity can be estimated by comparing the Rabi rates of different ions. Note that the oscillation for the left ion is not visible here due to the insufficient time points.
(e) Schematic of the four transverse (drumhead) modes with frequencies $\omega_k=2\pi\times (2.284,\,2.216,\,2.167,\,2.138)\,$MHz ($k=1,\,2,\,3,\,4$). The blue, gray and orange colors of the ions represent the positive, zero and negative mode amplitudes, respectively. The vertical dashed lines indicate the laser detuning for two-qubit gates between the left-right (LR), up-down (UD) and left-up (LU) ion pairs.
} \label{fig1}
\end{figure}

\emph{Experimental setup.}
Our experimental setup is sketched in Fig.~\ref{fig1}(a). We design a special configuration of the blade trap, as described in Supplementary Information, to obtain a 2D crystal of ${}^{171}\mathrm{Yb}^+$ ions perpendicular to the imaging direction while allowing wide optical access. A DC bias voltage is applied on both RF electrodes to split the radial trap frequency into $\omega_x=2\pi\times 0.803\,$MHz and $\omega_y=2\pi\times 2.284\,$MHz. We further use DC electrodes for the axial trapping $\omega_z=2\pi\times 0.553\,$MHz and obtain a 2D crystal whose major axis is oriented in the $z$ direction as shown in Fig.~\ref{fig1}(b). In the following we label our four trapped ions in a 2D crystal by left (L), right (R), up (U) and down (D). Due to the overlap between the images of adjacent ions and the low photon collection efficiency, there exists considerable crosstalk error in the single-shot measurement of the four qubits encoded in $\ket{0}\equiv \ket{S_{1/2},F=0,m_F=0}$ and $\ket{1}\equiv \ket{S_{1/2},F=1,m_F=0}$ under $370\,$nm detection laser. In the future, this can be improved by electron shelving to the $D_{5/2}$ and $F_{7/2}$ levels \cite{Roman2020,edmunds2020scalable}, but in this experiment for the purpose of calibrating individual addressing and two-qubit gate errors, it suffices to recover the population in the $2^4=16$ computational basis states by the maximum likelihood method. More details can be found in Supplementary Information.

We use two pairs of symmetrically placed crossed AODs for individual addressing of the $355\,$nm Raman laser beams \cite{patent-ZL202110047218.2}. By tuning the driving frequencies on the horizontal or vertical AODs, the addressing beam can be scanned along the $z$ or $x$ directions, respectively.
The objectives for the two addressing beams have the same focal length so that the two pairs of crossed AODs have equal driving frequencies when addressing the same target ion. In this way, the frequency shift introduced by the AODs can be cancelled in the Raman transition \cite{patent-ZL202110047218.2}.
Each Raman beam has a waist radius (where the intensity drops to $1/e^2$) of about $1.5\,\mu$m as compared with the distance between adjacent ions of about $5\,\mu$m.
To estimate the crosstalk for individual addressing, we drive a resonant Raman transition between $\ket{0}$ and $\ket{1}$ on one target ion, and measure the Rabi oscillation of all the ions. An example is shown in Fig.~\ref{fig1}(c) and (d) for the addressing beam on the left ion. By fitting the Rabi frequency of $\Omega_L=2\pi\times 1.04\,$MHz for the left ion, and $\Omega_U=\Omega_D=2\pi\times 2.7\,$kHz for the up and down ions (the Rabi rate for the right ion is even smaller due to its larger distance), we estimate the crosstalk infidelity for a single-qubit $\pi$ pulse to be $[(\pi/2)(\Omega_D/\Omega_L)]^2=2\times 10^{-5}$. Similarly, we measure the crosstalk for other target ions as shown in Fig.~S3 of Supplementary Information and obtain a maximal crosstalk infidelity of $0.08\%$.

The four transverse (drumhead) phonon modes of the four-ion crystal are illustrated in Fig.~\ref{fig1}(e). Mediated by these phonon modes, two-qubit entangling gates between any ion pair, say, LR, UD or LU, can be realized by spin-dependent forces via, e.g., a phase-modulated gate sequence \cite{greenPhaseModulatedDecouplingError2015}.

\begin{figure}[!tbp]
	\centering
	\includegraphics[width=\linewidth]{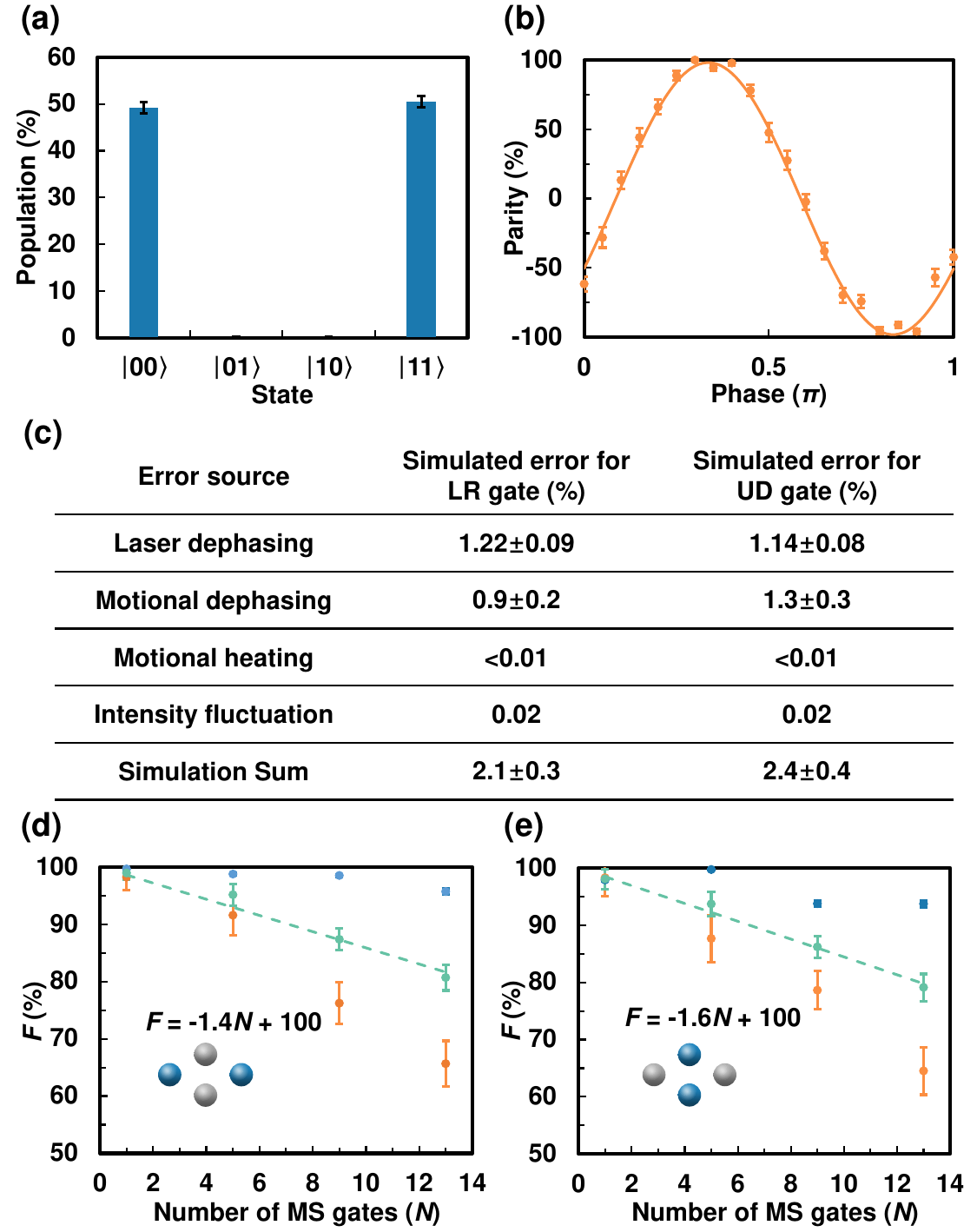}
	\caption {Entangling gates between ion pairs in the same row or the same column. (a) Population for the LR ion pair after an entangling gate. (b) Parity oscillation for the LR ion pair after an entangling gate. (c) Theoretical estimation of gate errors for the LR and the UD ion pairs. The laser dephasing, motional dephasing and motional heating effects are modeled by Lindblad operators. The intensity fluctuation is described by a shot-to-shot variation. (d) Experimentally measured Bell-state fidelity for the LR ion pair vs. the number of entangling gates. The blue and the orange data points represent the population and parity, respectively, while the fidelity is their average as the green dots. By fitting the slope of the fidelity, we extract a gate infidelity of about $1.4(2)\%$. (e) Similar plot for the UD ion pair with an extracted gate infidelity of about $1.6(1)\%$. All the error bars represent one standard deviation.} \label{fig2}
\end{figure}

\emph{Individually addressed two-qubit gates.}
For universal quantum computing, we want two-qubit entangling gates between all ion pairs, or at least a connected graph of them. For our four-ion crystal, this set includes a horizontal pair (LR), a vertical pair (UD), and four diagonal pairs (LU, LD, RU and RD). Here we demonstrate the entangling gates for the LR, UD and LU pairs, and expect similar performance for other diagonal pairs by symmetry.

With the crossed AODs, the horizontal or the vertical pair can be easily addressed by applying two frequency components to the horizontal or the vertical AODs, respectively. Then we can use bichromatic $355\,$nm Raman laser to generate spin-dependent forces on the target ions \cite{debnath2016programmable} and use a phase-modulated gate sequence to disentangle the spin and the motional states at the end of the gate \cite{greenPhaseModulatedDecouplingError2015}. For the horizontal pair, as shown in Fig.~\ref{fig1}(e), the fourth phonon mode has zero mode coefficient and thus can be ignored in the gate design. We set the Raman laser detuning $\mu=(\omega_2+\omega_3)/2$ in the middle of Mode 2 and Mode 3 such that after each segment of $t=4\pi/(\omega_2-\omega_3)$ (one loop in the phase space) they will be disentangled simultaneously. We further choose the phase shift between two segments to decouple the center-of-mass (COM) mode and obtain the total gate time $T=2t=81.6\,\mu$s. Finally, we scale the laser intensity for the accumulated two-spin phase to give us a maximally entangled two-qubit gate. As shown in Figs.~\ref{fig2}(a) and (b), we obtain a population of $99.6(1)\%$ in $\ket{00}$ and $\ket{11}$, and a parity contrast of $98(2)\%$, thus a Bell state fidelity of $99(1)\%$ for the LR pair \cite{debnath2016programmable}. As we show in Fig.~\ref{fig2}(c), the error mainly comes from the laser dephasing time of $4\,$ms due to the optical phase fluctuation between the two Raman beams, and the motional dephasing time of $3\,$ms. However, note that these theoretical analyses are based on a white-noise model, which may over-estimate the gate error as the actual noise can be dominated by the low-frequency part. More details can be found in Supplementary Information.

Similarly, we can address the UD ion pair by two frequency components on the vertical AODs. These two ions do not participate in the second mode, and we set the Raman laser detuning as $\mu=2\omega_4-\omega_3$ such that after each segment of $t=2\pi/(\omega_3-\omega_4)$, one loop in the phase space will be traversed by the fourth mode while two by the third mode. Again we use two segments with an adjustable phase shift to disentangle the COM mode. At the total gate time $T=2t=69.0\,\mu$s, we obtain a Bell state fidelity of $98(2)\%$.

In Figs.~\ref{fig2}(d) and (e), we repeat the two-qubit gates for an odd number of times and examine the decay of the Bell state fidelity vs. the repetition number. Assuming a dominant incoherent error, this can be used to separate the gate infidelity from the state-preparation-and-measurement (SPAM) error \cite{PhysRevLett.125.150505,PhysRevA.107.032617}, although in our case the detection error is already corrected reasonably well and indeed we fit an intercept close to $100\%$. Nevertheless, it still allows us to reduce the statistical uncertainty and we fit the gate infidelities to be $1.4(2)\%$ and $1.6(1)\%$, respectively, for LR and UD pairs.

\begin{figure}[!tbp]
	\centering
	\includegraphics[width=\linewidth]{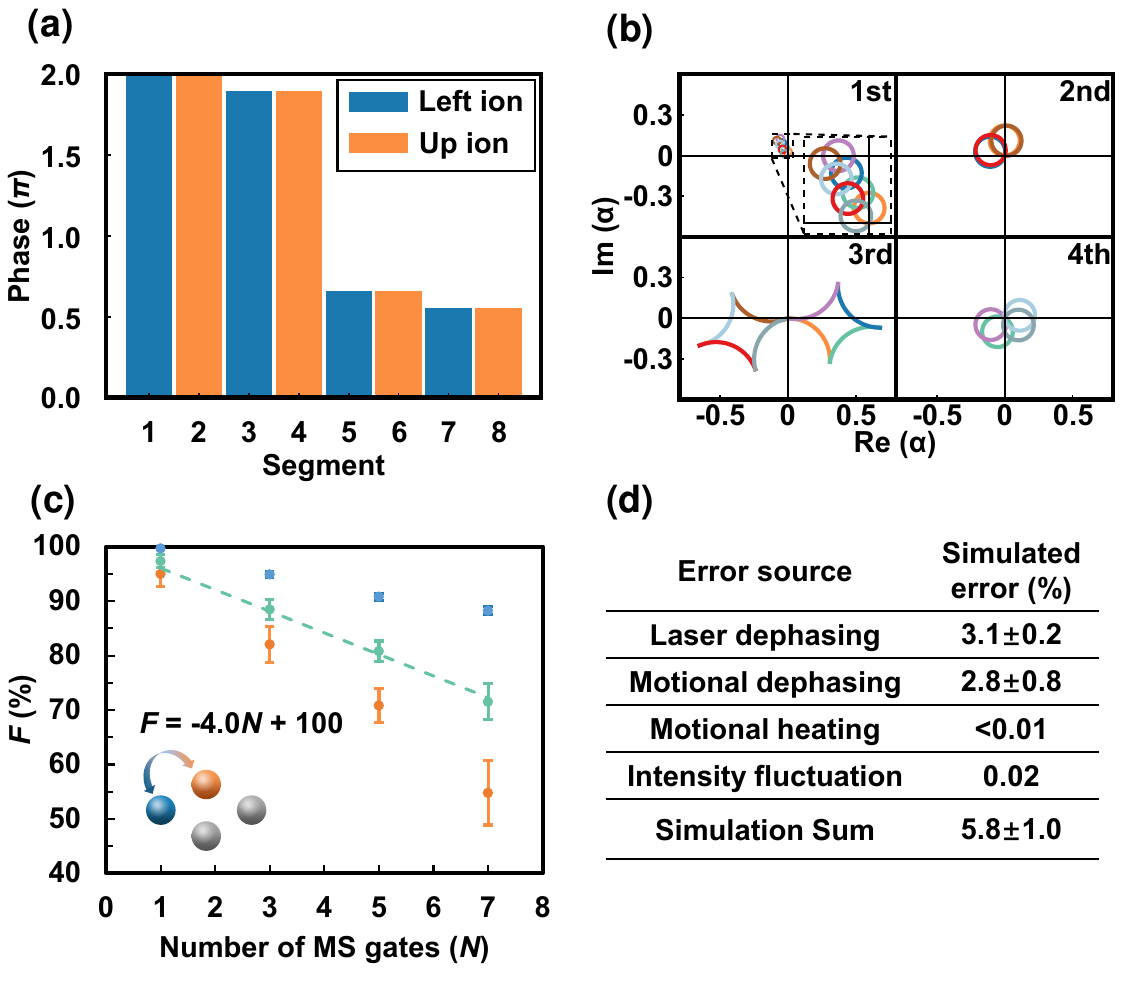}
	\caption {Entangling gate between a diagonal ion pair. (a) Phase modulation sequence for the LU ion pair. We alternatingly apply the driving laser on the two target ions, indicated by the color of the bars, with a fixed laser detuning $\mu_{LU}=(\omega_2+\omega_4)/2$. Each segment takes time $t=4\pi/(\omega_2-\omega_4)$ with the laser phase given by the height of the bars. We set a separation $\Delta t=2\,\mu$s between adjacent segments to turn the driving laser off and on, so as to avoid crosstalk during the switching. (b) Theoretical phase space trajectories for the four phonon modes when the spin state is $\ket{++}$. Different segments are indicated by different colors. (c) Similar plot as Figs.~\ref{fig2}(d) and (e) for the LU ion pair. A gate infidelity of $4.0(3)\%$ is fitted. (d) Similar plot as Fig.~\ref{fig2}(c) for theoretically simulated gate errors.} \label{fig3}
\end{figure}

To address the LU ion pair by the crossed AODs is more complicated. If we apply two frequency components to both the horizontal and the vertical AODs, four light spots in a rectangular pattern will be generated. It not only decreases the laser intensity on the target ions, but also causes stronger crosstalk on the other ions due to the undesired light spots. To circumvent this problem, we develop an alternating gate sequence such that at any time only a single ion will be addressed. Note that although we demonstrate this scheme on crossed AODs, it is also compatible with other 2D addressing techniques.

Similar to the commonly used Molmer-Sorensen gate with amplitude, phase or frequency modulation \cite{Zhu2006,greenPhaseModulatedDecouplingError2015,PhysRevA.98.032318}, our alternating gate sequence also aims to disentangle the spin and the phonon modes while accumulating a desired two-qubit phase for maximal entanglement. Note that the two-qubit phase comes from the commutation relation between a new displacement in the phase space and an accumulative one (see Supplementary Information), and does not require simultaneous addressing of both ions. Specifically, for the LU ion pair we design a phase modulation sequence as shown in Fig.~\ref{fig3}(a) with a Raman laser detuning $\mu=(\omega_2+\omega_4)/2$ in the middle of Mode $2$ and Mode $4$, and a segment duration $t=4\pi/(\omega_2-\omega_4)$. Note that these two modes have participation of only one ion from the desired LU ion pair, hence do not contribute to the two-qubit phase. Therefore we use the above elementary segment to disentangle them from the spin states trivially. The rest two phonon modes can be disentangled by four phase-modulation segments \cite{greenPhaseModulatedDecouplingError2015}, and we apply such a sequence alternatingly on the two target ions, obtaining the sequence in Fig.~\ref{fig3}(a). We set a separation $\Delta t=2\,\mu$s between adjacent segments to avoid crosstalk when switching the addressing beam. This gives us a total gate time of $T=8t+7\Delta t=219.1\,\mu$s. The corresponding phase space trajectories for the four phonon modes from the initial spin state $\ket{++}$ (an eigenstate of the laser-induced spin-dependent force, see Supplementary Information) is shown in Fig.~\ref{fig3}(b), with different segments indicated by different colors. As designed, all these trajectories close at the end of the gate. Then we repeat the entangling gate for an odd number of times in Fig.~\ref{fig3}(c) and fit a gate infidelity of $4.0(3)\%$, which is again dominated by the laser dephasing and motional dephasing effects as shown in Fig.~\ref{fig3}(d).

Note that the above construction of gate sequences to disentangle all the phonon modes exactly can become inefficient as the ion number increases \cite{greenPhaseModulatedDecouplingError2015}. As we describe in Supplementary Information, we can also use a moderate number of segments to decouple the spin and the phonon modes approximately while still achieving high gate fidelity.

\begin{figure}[!tbp]
	\centering
	\includegraphics[width=\linewidth]{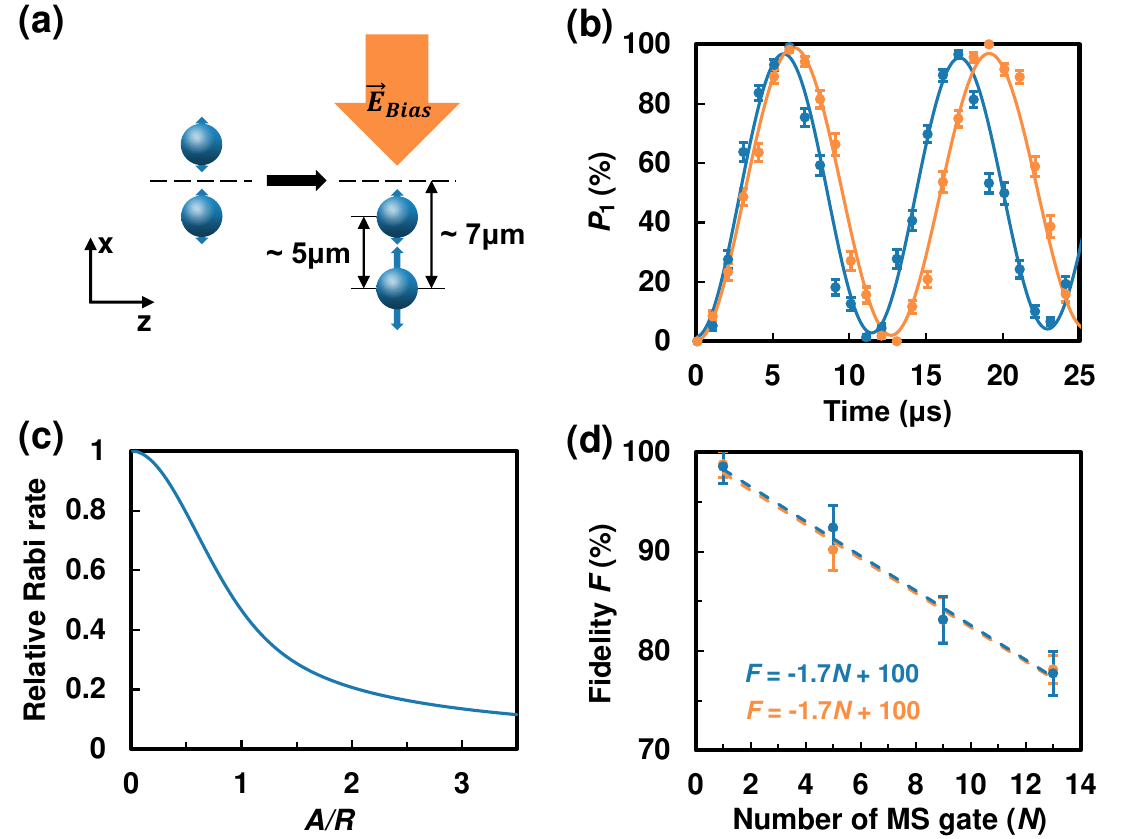}
	\caption {Entangling gates under micromotion. (a) We orient an ion pair along the $x$ axis with a separation of about $5\,\mu$m. Then we push the ion pair away from the micromotion-free $z$ axis by a bias electric field. The relative micromotion amplitudes of the two ions are indicated by the blue arrows, which are not to scale with the ion spacing or their displacements. (b) Rabi oscillations of the lower ion under a Raman laser with the same intensity before (blue) and after (orange) the displacement. (c) Theoretical relative Rabi rate $r$ vs. the micromotion amplitude $A$ in the unit of the laser waist radius $R=1.5\,\mu$m. (d) Bell-state fidelity vs. the number of entangling gates for the ion pairs before (blue) and after (orange) the displacement. We fit both gate infidelities to be $1.7(1)\%$. All the error bars represent one standard deviation.} \label{fig4}
\end{figure}

\emph{Effects of Micromotion.}
In our four-ion crystal, the LR pair locates on the RF null axis, while the UD pair has nonzero excess micromotion proportional to the deviation from the RF null. From the above results we see that the entangling gates on these two ion pairs have similar infidelities and that their difference is mainly caused by the different gate time and phonon excitations rather than the existence of micromotion. This suggests that micromotion of ions in a 2D crystal is not a limiting factor for the gate performance.

We further examine the influence of micromotion in Fig.~\ref{fig4} using a two-ion crystal aligned in the $x$ direction. We apply a bias voltage on one DC and one RF electrode to create a bias electric field along the $x$ direction and push the ion pair away from the RF null axis. Theoretically, the micromotion amplitude is given by $A=qx/2$ where $x$ is the coordinate of the concerned ion and $q=0.12$ is the Mathieu parameter in that direction \cite{leibfriedQuantumDynamicsSingle2003}. As the micromotion amplitude becomes comparable to the radius $R=1.5\,\mu$m of the addressing beams, we expect a reduction in the effective laser intensity felt by the ions. This can be seen from Fig.~\ref{fig4}(b) where we plot the Raman Rabi oscillation of the lower ion before (blue) and after (orange) the displacement under the same laser intensity. We fit the corresponding Rabi frequencies and get their ratio $\Omega_2/\Omega_1=0.90$. If we assume a Gaussian profile of the addressing beams, we can theoretically compute the relative Rabi rate [Fig.~\ref{fig4}(c)] $r(A)=(1/2\pi)\int_0^{2\pi}e^{-2(A\cos\theta/R)^2}d\theta=e^{-A^2/R^2}I_0(A^2/R^2)$ where $I_0(x)$ is the modified Bessel function of the first kind. Here we average over an RF period without worrying about its initial phase because the RF frequency $\omega_{\mathrm{rf}}=2\pi\times 37\,$MHz is much higher than any other timescales in this experiment. This calculation predicts a Rabi rate ratio of $0.935$. Small deviation between theoretical and experimental results may come from the non-Gaussian distribution of the laser intensity.

We further perform two-qubit gates on the ion pair before and after the displacement. We set the laser detuning to the middle of the two phonon modes and evolve both modes by four loops in the phase space for a gate time of about $100\,\mu$s. As shown in Fig.~\ref{fig4}(d), after calibrating the laser intensity, we get almost the same gate performance with small (blue) or large (orange) excess micromotion.
Also note that due to the different micromotion amplitudes of the two ions, their effective Rabi rates will differ under the same laser intensity. Ideally, a two-qubit phase can still be obtained which is proportional to the product of the two Rabi frequencies, but it may increase the spontaneous emission error from the excited states. Therefore, here we calibrate the intensities of the two addressing beams to ensure the same effective Rabi frequencies on the two target ions.

In this experiment, we only push the ions to about $7\,\mu$m from the RF null with a micromotion amplitude of about $420\,$nm. This is restricted by the stability of the crystal under large excess micromotion. Using a cryogenic ion trap, much larger 2D ion crystals can be stably trapped \cite{guo2023siteresolved}, and we expect the ultimate limit for 2D individual addressing to be when the micromotion amplitude $A$ becomes comparable to the ion spacing $d\approx 5\,\mu$m. According to Fig.~\ref{fig4}(c), in such cases the required laser intensity may be increased by about eight times to compensate the micromotion effect. Nevertheless, this does not restrict the gate fidelity because all the other noise sources like the spontaneous emission are rescaled as well.

To sum up, we demonstrate the individual addressing of a 2D ion crystal by crossed AODs and achieve all-to-all connected two-qubit gates. For diagonal ion pairs, we develop an alternating gate sequence such that at most one ion needs to be addressed at any time. We further show that the gate fidelities on a 2D crystal is not limited by the micromotion which can be compensated by calibrating the intensity of the addressing beams.
Our current two-qubit gate fidelities are restricted by the short laser and motional dephasing time, and can be improved by shortening and stabilizing the optical path and by locking the trap frequency in future upgrades.

\begin{acknowledgments}
This work was supported by Innovation Program for Quantum Science and Technology (2021ZD0301601), Tsinghua University Initiative Scientific Research Program, and the Ministry of Education of China. L.M.D. acknowledges in addition support from the New Cornerstone Science Foundation through the New Cornerstone Investigator Program. Y.K.W. acknowledges in addition support from Tsinghua University Dushi program. C.Z. acknowledges in addition support from the Shui-Mu Scholar postdoctoral fellowship from Tsinghua University. P.Y.H. acknowledges the start-up fund from Tsinghua University.
\end{acknowledgments}


\begin{thebibliography}{49}%
\makeatletter
\providecommand \@ifxundefined [1]{%
 \@ifx{#1\undefined}
}%
\providecommand \@ifnum [1]{%
 \ifnum #1\expandafter \@firstoftwo
 \else \expandafter \@secondoftwo
 \fi
}%
\providecommand \@ifx [1]{%
 \ifx #1\expandafter \@firstoftwo
 \else \expandafter \@secondoftwo
 \fi
}%
\providecommand \natexlab [1]{#1}%
\providecommand \enquote  [1]{``#1''}%
\providecommand \bibnamefont  [1]{#1}%
\providecommand \bibfnamefont [1]{#1}%
\providecommand \citenamefont [1]{#1}%
\providecommand \href@noop [0]{\@secondoftwo}%
\providecommand \href [0]{\begingroup \@sanitize@url \@href}%
\providecommand \@href[1]{\@@startlink{#1}\@@href}%
\providecommand \@@href[1]{\endgroup#1\@@endlink}%
\providecommand \@sanitize@url [0]{\catcode `\\12\catcode `\$12\catcode
  `\&12\catcode `\#12\catcode `\^12\catcode `\_12\catcode `\%12\relax}%
\providecommand \@@startlink[1]{}%
\providecommand \@@endlink[0]{}%
\providecommand \url  [0]{\begingroup\@sanitize@url \@url }%
\providecommand \@url [1]{\endgroup\@href {#1}{\urlprefix }}%
\providecommand \urlprefix  [0]{URL }%
\providecommand \Eprint [0]{\href }%
\providecommand \doibase [0]{http://dx.doi.org/}%
\providecommand \selectlanguage [0]{\@gobble}%
\providecommand \bibinfo  [0]{\@secondoftwo}%
\providecommand \bibfield  [0]{\@secondoftwo}%
\providecommand \translation [1]{[#1]}%
\providecommand \BibitemOpen [0]{}%
\providecommand \bibitemStop [0]{}%
\providecommand \bibitemNoStop [0]{.\EOS\space}%
\providecommand \EOS [0]{\spacefactor3000\relax}%
\providecommand \BibitemShut  [1]{\csname bibitem#1\endcsname}%
\let\auto@bib@innerbib\@empty
\bibitem [{\citenamefont {Nielsen}\ and\ \citenamefont
  {Chuang}(2000)}]{nielsen2000quantum}%
  \BibitemOpen
  \bibfield  {author} {\bibinfo {author} {\bibfnamefont {M.}~\bibnamefont
  {Nielsen}}\ and\ \bibinfo {author} {\bibfnamefont {I.}~\bibnamefont
  {Chuang}},\ }\href@noop {} {\emph {\bibinfo {title} {Quantum Computation and
  Quantum Information}}}\ (\bibinfo  {publisher} {Cambridge University Press},\
  \bibinfo {year} {2000})\BibitemShut {NoStop}%
\bibitem [{\citenamefont {Acharya}\ \emph {et~al.}(2023)\citenamefont
  {Acharya}, \citenamefont {Aleiner}, \citenamefont {Allen}, \citenamefont
  {Andersen}, \citenamefont {Ansmann}, \citenamefont {Arute}, \citenamefont
  {Arya}, \citenamefont {Asfaw}, \citenamefont {Atalaya}, \citenamefont
  {Babbush}, \citenamefont {Bacon}, \citenamefont {Bardin}, \citenamefont
  {Basso}, \citenamefont {Bengtsson}, \citenamefont {Boixo}, \citenamefont
  {Bortoli}, \citenamefont {Bourassa}, \citenamefont {Bovaird}, \citenamefont
  {Brill}, \citenamefont {Broughton}, \citenamefont {Buckley}, \citenamefont
  {Buell}, \citenamefont {Burger}, \citenamefont {Burkett}, \citenamefont
  {Bushnell}, \citenamefont {Chen}, \citenamefont {Chen}, \citenamefont
  {Chiaro}, \citenamefont {Cogan}, \citenamefont {Collins}, \citenamefont
  {Conner}, \citenamefont {Courtney}, \citenamefont {Crook}, \citenamefont
  {Curtin}, \citenamefont {Debroy}, \citenamefont {Del Toro~Barba},
  \citenamefont {Demura}, \citenamefont {Dunsworth}, \citenamefont {Eppens},
  \citenamefont {Erickson}, \citenamefont {Faoro}, \citenamefont {Farhi},
  \citenamefont {Fatemi}, \citenamefont {Flores~Burgos}, \citenamefont
  {Forati}, \citenamefont {Fowler}, \citenamefont {Foxen}, \citenamefont
  {Giang}, \citenamefont {Gidney}, \citenamefont {Gilboa}, \citenamefont
  {Giustina}, \citenamefont {Grajales~Dau}, \citenamefont {Gross},
  \citenamefont {Habegger}, \citenamefont {Hamilton}, \citenamefont {Harrigan},
  \citenamefont {Harrington}, \citenamefont {Higgott}, \citenamefont {Hilton},
  \citenamefont {Hoffmann}, \citenamefont {Hong}, \citenamefont {Huang},
  \citenamefont {Huff}, \citenamefont {Huggins}, \citenamefont {Ioffe},
  \citenamefont {Isakov}, \citenamefont {Iveland}, \citenamefont {Jeffrey},
  \citenamefont {Jiang}, \citenamefont {Jones}, \citenamefont {Juhas},
  \citenamefont {Kafri}, \citenamefont {Kechedzhi}, \citenamefont {Kelly},
  \citenamefont {Khattar}, \citenamefont {Khezri}, \citenamefont
  {Kieferov{\'a}}, \citenamefont {Kim}, \citenamefont {Kitaev}, \citenamefont
  {Klimov}, \citenamefont {Klots}, \citenamefont {Korotkov}, \citenamefont
  {Kostritsa}, \citenamefont {Kreikebaum}, \citenamefont {Landhuis},
  \citenamefont {Laptev}, \citenamefont {Lau}, \citenamefont {Laws},
  \citenamefont {Lee}, \citenamefont {Lee}, \citenamefont {Lester},
  \citenamefont {Lill}, \citenamefont {Liu}, \citenamefont {Locharla},
  \citenamefont {Lucero}, \citenamefont {Malone}, \citenamefont {Marshall},
  \citenamefont {Martin}, \citenamefont {McClean}, \citenamefont {McCourt},
  \citenamefont {McEwen}, \citenamefont {Megrant}, \citenamefont
  {Meurer~Costa}, \citenamefont {Mi}, \citenamefont {Miao}, \citenamefont
  {Mohseni}, \citenamefont {Montazeri}, \citenamefont {Morvan}, \citenamefont
  {Mount}, \citenamefont {Mruczkiewicz}, \citenamefont {Naaman}, \citenamefont
  {Neeley}, \citenamefont {Neill}, \citenamefont {Nersisyan}, \citenamefont
  {Neven}, \citenamefont {Newman}, \citenamefont {Ng}, \citenamefont {Nguyen},
  \citenamefont {Nguyen}, \citenamefont {Niu}, \citenamefont {O'Brien},
  \citenamefont {Opremcak}, \citenamefont {Platt}, \citenamefont {Petukhov},
  \citenamefont {Potter}, \citenamefont {Pryadko}, \citenamefont {Quintana},
  \citenamefont {Roushan}, \citenamefont {Rubin}, \citenamefont {Saei},
  \citenamefont {Sank}, \citenamefont {Sankaragomathi}, \citenamefont
  {Satzinger}, \citenamefont {Schurkus}, \citenamefont {Schuster},
  \citenamefont {Shearn}, \citenamefont {Shorter}, \citenamefont {Shvarts},
  \citenamefont {Skruzny}, \citenamefont {Smelyanskiy}, \citenamefont {Smith},
  \citenamefont {Sterling}, \citenamefont {Strain}, \citenamefont {Szalay},
  \citenamefont {Torres}, \citenamefont {Vidal}, \citenamefont {Villalonga},
  \citenamefont {Vollgraff~Heidweiller}, \citenamefont {White}, \citenamefont
  {Xing}, \citenamefont {Yao}, \citenamefont {Yeh}, \citenamefont {Yoo},
  \citenamefont {Young}, \citenamefont {Zalcman}, \citenamefont {Zhang},
  \citenamefont {Zhu},\ and\ \citenamefont {AI}}]{Acharya2023}%
  \BibitemOpen
  \bibfield  {author} {\bibinfo {author} {\bibfnamefont {R.}~\bibnamefont
  {Acharya}}, \bibinfo {author} {\bibfnamefont {I.}~\bibnamefont {Aleiner}},
  \bibinfo {author} {\bibfnamefont {R.}~\bibnamefont {Allen}}, \bibinfo
  {author} {\bibfnamefont {T.~I.}\ \bibnamefont {Andersen}}, \bibinfo {author}
  {\bibfnamefont {M.}~\bibnamefont {Ansmann}}, \bibinfo {author} {\bibfnamefont
  {F.}~\bibnamefont {Arute}}, \bibinfo {author} {\bibfnamefont
  {K.}~\bibnamefont {Arya}}, \bibinfo {author} {\bibfnamefont {A.}~\bibnamefont
  {Asfaw}}, \bibinfo {author} {\bibfnamefont {J.}~\bibnamefont {Atalaya}},
  \bibinfo {author} {\bibfnamefont {R.}~\bibnamefont {Babbush}}, \bibinfo
  {author} {\bibfnamefont {D.}~\bibnamefont {Bacon}}, \bibinfo {author}
  {\bibfnamefont {J.~C.}\ \bibnamefont {Bardin}}, \bibinfo {author}
  {\bibfnamefont {J.}~\bibnamefont {Basso}}, \bibinfo {author} {\bibfnamefont
  {A.}~\bibnamefont {Bengtsson}}, \bibinfo {author} {\bibfnamefont
  {S.}~\bibnamefont {Boixo}}, \bibinfo {author} {\bibfnamefont
  {G.}~\bibnamefont {Bortoli}}, \bibinfo {author} {\bibfnamefont
  {A.}~\bibnamefont {Bourassa}}, \bibinfo {author} {\bibfnamefont
  {J.}~\bibnamefont {Bovaird}}, \bibinfo {author} {\bibfnamefont
  {L.}~\bibnamefont {Brill}}, \bibinfo {author} {\bibfnamefont
  {M.}~\bibnamefont {Broughton}}, \bibinfo {author} {\bibfnamefont {B.~B.}\
  \bibnamefont {Buckley}}, \bibinfo {author} {\bibfnamefont {D.~A.}\
  \bibnamefont {Buell}}, \bibinfo {author} {\bibfnamefont {T.}~\bibnamefont
  {Burger}}, \bibinfo {author} {\bibfnamefont {B.}~\bibnamefont {Burkett}},
  \bibinfo {author} {\bibfnamefont {N.}~\bibnamefont {Bushnell}}, \bibinfo
  {author} {\bibfnamefont {Y.}~\bibnamefont {Chen}}, \bibinfo {author}
  {\bibfnamefont {Z.}~\bibnamefont {Chen}}, \bibinfo {author} {\bibfnamefont
  {B.}~\bibnamefont {Chiaro}}, \bibinfo {author} {\bibfnamefont
  {J.}~\bibnamefont {Cogan}}, \bibinfo {author} {\bibfnamefont
  {R.}~\bibnamefont {Collins}}, \bibinfo {author} {\bibfnamefont
  {P.}~\bibnamefont {Conner}}, \bibinfo {author} {\bibfnamefont
  {W.}~\bibnamefont {Courtney}}, \bibinfo {author} {\bibfnamefont {A.~L.}\
  \bibnamefont {Crook}}, \bibinfo {author} {\bibfnamefont {B.}~\bibnamefont
  {Curtin}}, \bibinfo {author} {\bibfnamefont {D.~M.}\ \bibnamefont {Debroy}},
  \bibinfo {author} {\bibfnamefont {A.}~\bibnamefont {Del Toro~Barba}},
  \bibinfo {author} {\bibfnamefont {S.}~\bibnamefont {Demura}}, \bibinfo
  {author} {\bibfnamefont {A.}~\bibnamefont {Dunsworth}}, \bibinfo {author}
  {\bibfnamefont {D.}~\bibnamefont {Eppens}}, \bibinfo {author} {\bibfnamefont
  {C.}~\bibnamefont {Erickson}}, \bibinfo {author} {\bibfnamefont
  {L.}~\bibnamefont {Faoro}}, \bibinfo {author} {\bibfnamefont
  {E.}~\bibnamefont {Farhi}}, \bibinfo {author} {\bibfnamefont
  {R.}~\bibnamefont {Fatemi}}, \bibinfo {author} {\bibfnamefont
  {L.}~\bibnamefont {Flores~Burgos}}, \bibinfo {author} {\bibfnamefont
  {E.}~\bibnamefont {Forati}}, \bibinfo {author} {\bibfnamefont {A.~G.}\
  \bibnamefont {Fowler}}, \bibinfo {author} {\bibfnamefont {B.}~\bibnamefont
  {Foxen}}, \bibinfo {author} {\bibfnamefont {W.}~\bibnamefont {Giang}},
  \bibinfo {author} {\bibfnamefont {C.}~\bibnamefont {Gidney}}, \bibinfo
  {author} {\bibfnamefont {D.}~\bibnamefont {Gilboa}}, \bibinfo {author}
  {\bibfnamefont {M.}~\bibnamefont {Giustina}}, \bibinfo {author}
  {\bibfnamefont {A.}~\bibnamefont {Grajales~Dau}}, \bibinfo {author}
  {\bibfnamefont {J.~A.}\ \bibnamefont {Gross}}, \bibinfo {author}
  {\bibfnamefont {S.}~\bibnamefont {Habegger}}, \bibinfo {author}
  {\bibfnamefont {M.~C.}\ \bibnamefont {Hamilton}}, \bibinfo {author}
  {\bibfnamefont {M.~P.}\ \bibnamefont {Harrigan}}, \bibinfo {author}
  {\bibfnamefont {S.~D.}\ \bibnamefont {Harrington}}, \bibinfo {author}
  {\bibfnamefont {O.}~\bibnamefont {Higgott}}, \bibinfo {author} {\bibfnamefont
  {J.}~\bibnamefont {Hilton}}, \bibinfo {author} {\bibfnamefont
  {M.}~\bibnamefont {Hoffmann}}, \bibinfo {author} {\bibfnamefont
  {S.}~\bibnamefont {Hong}}, \bibinfo {author} {\bibfnamefont {T.}~\bibnamefont
  {Huang}}, \bibinfo {author} {\bibfnamefont {A.}~\bibnamefont {Huff}},
  \bibinfo {author} {\bibfnamefont {W.~J.}\ \bibnamefont {Huggins}}, \bibinfo
  {author} {\bibfnamefont {L.~B.}\ \bibnamefont {Ioffe}}, \bibinfo {author}
  {\bibfnamefont {S.~V.}\ \bibnamefont {Isakov}}, \bibinfo {author}
  {\bibfnamefont {J.}~\bibnamefont {Iveland}}, \bibinfo {author} {\bibfnamefont
  {E.}~\bibnamefont {Jeffrey}}, \bibinfo {author} {\bibfnamefont
  {Z.}~\bibnamefont {Jiang}}, \bibinfo {author} {\bibfnamefont
  {C.}~\bibnamefont {Jones}}, \bibinfo {author} {\bibfnamefont
  {P.}~\bibnamefont {Juhas}}, \bibinfo {author} {\bibfnamefont
  {D.}~\bibnamefont {Kafri}}, \bibinfo {author} {\bibfnamefont
  {K.}~\bibnamefont {Kechedzhi}}, \bibinfo {author} {\bibfnamefont
  {J.}~\bibnamefont {Kelly}}, \bibinfo {author} {\bibfnamefont
  {T.}~\bibnamefont {Khattar}}, \bibinfo {author} {\bibfnamefont
  {M.}~\bibnamefont {Khezri}}, \bibinfo {author} {\bibfnamefont
  {M.}~\bibnamefont {Kieferov{\'a}}}, \bibinfo {author} {\bibfnamefont
  {S.}~\bibnamefont {Kim}}, \bibinfo {author} {\bibfnamefont {A.}~\bibnamefont
  {Kitaev}}, \bibinfo {author} {\bibfnamefont {P.~V.}\ \bibnamefont {Klimov}},
  \bibinfo {author} {\bibfnamefont {A.~R.}\ \bibnamefont {Klots}}, \bibinfo
  {author} {\bibfnamefont {A.~N.}\ \bibnamefont {Korotkov}}, \bibinfo {author}
  {\bibfnamefont {F.}~\bibnamefont {Kostritsa}}, \bibinfo {author}
  {\bibfnamefont {J.~M.}\ \bibnamefont {Kreikebaum}}, \bibinfo {author}
  {\bibfnamefont {D.}~\bibnamefont {Landhuis}}, \bibinfo {author}
  {\bibfnamefont {P.}~\bibnamefont {Laptev}}, \bibinfo {author} {\bibfnamefont
  {K.-M.}\ \bibnamefont {Lau}}, \bibinfo {author} {\bibfnamefont
  {L.}~\bibnamefont {Laws}}, \bibinfo {author} {\bibfnamefont {J.}~\bibnamefont
  {Lee}}, \bibinfo {author} {\bibfnamefont {K.}~\bibnamefont {Lee}}, \bibinfo
  {author} {\bibfnamefont {B.~J.}\ \bibnamefont {Lester}}, \bibinfo {author}
  {\bibfnamefont {A.}~\bibnamefont {Lill}}, \bibinfo {author} {\bibfnamefont
  {W.}~\bibnamefont {Liu}}, \bibinfo {author} {\bibfnamefont {A.}~\bibnamefont
  {Locharla}}, \bibinfo {author} {\bibfnamefont {E.}~\bibnamefont {Lucero}},
  \bibinfo {author} {\bibfnamefont {F.~D.}\ \bibnamefont {Malone}}, \bibinfo
  {author} {\bibfnamefont {J.}~\bibnamefont {Marshall}}, \bibinfo {author}
  {\bibfnamefont {O.}~\bibnamefont {Martin}}, \bibinfo {author} {\bibfnamefont
  {J.~R.}\ \bibnamefont {McClean}}, \bibinfo {author} {\bibfnamefont
  {T.}~\bibnamefont {McCourt}}, \bibinfo {author} {\bibfnamefont
  {M.}~\bibnamefont {McEwen}}, \bibinfo {author} {\bibfnamefont
  {A.}~\bibnamefont {Megrant}}, \bibinfo {author} {\bibfnamefont
  {B.}~\bibnamefont {Meurer~Costa}}, \bibinfo {author} {\bibfnamefont
  {X.}~\bibnamefont {Mi}}, \bibinfo {author} {\bibfnamefont {K.~C.}\
  \bibnamefont {Miao}}, \bibinfo {author} {\bibfnamefont {M.}~\bibnamefont
  {Mohseni}}, \bibinfo {author} {\bibfnamefont {S.}~\bibnamefont {Montazeri}},
  \bibinfo {author} {\bibfnamefont {A.}~\bibnamefont {Morvan}}, \bibinfo
  {author} {\bibfnamefont {E.}~\bibnamefont {Mount}}, \bibinfo {author}
  {\bibfnamefont {W.}~\bibnamefont {Mruczkiewicz}}, \bibinfo {author}
  {\bibfnamefont {O.}~\bibnamefont {Naaman}}, \bibinfo {author} {\bibfnamefont
  {M.}~\bibnamefont {Neeley}}, \bibinfo {author} {\bibfnamefont
  {C.}~\bibnamefont {Neill}}, \bibinfo {author} {\bibfnamefont
  {A.}~\bibnamefont {Nersisyan}}, \bibinfo {author} {\bibfnamefont
  {H.}~\bibnamefont {Neven}}, \bibinfo {author} {\bibfnamefont
  {M.}~\bibnamefont {Newman}}, \bibinfo {author} {\bibfnamefont {J.~H.}\
  \bibnamefont {Ng}}, \bibinfo {author} {\bibfnamefont {A.}~\bibnamefont
  {Nguyen}}, \bibinfo {author} {\bibfnamefont {M.}~\bibnamefont {Nguyen}},
  \bibinfo {author} {\bibfnamefont {M.~Y.}\ \bibnamefont {Niu}}, \bibinfo
  {author} {\bibfnamefont {T.~E.}\ \bibnamefont {O'Brien}}, \bibinfo {author}
  {\bibfnamefont {A.}~\bibnamefont {Opremcak}}, \bibinfo {author}
  {\bibfnamefont {J.}~\bibnamefont {Platt}}, \bibinfo {author} {\bibfnamefont
  {A.}~\bibnamefont {Petukhov}}, \bibinfo {author} {\bibfnamefont
  {R.}~\bibnamefont {Potter}}, \bibinfo {author} {\bibfnamefont {L.~P.}\
  \bibnamefont {Pryadko}}, \bibinfo {author} {\bibfnamefont {C.}~\bibnamefont
  {Quintana}}, \bibinfo {author} {\bibfnamefont {P.}~\bibnamefont {Roushan}},
  \bibinfo {author} {\bibfnamefont {N.~C.}\ \bibnamefont {Rubin}}, \bibinfo
  {author} {\bibfnamefont {N.}~\bibnamefont {Saei}}, \bibinfo {author}
  {\bibfnamefont {D.}~\bibnamefont {Sank}}, \bibinfo {author} {\bibfnamefont
  {K.}~\bibnamefont {Sankaragomathi}}, \bibinfo {author} {\bibfnamefont
  {K.~J.}\ \bibnamefont {Satzinger}}, \bibinfo {author} {\bibfnamefont {H.~F.}\
  \bibnamefont {Schurkus}}, \bibinfo {author} {\bibfnamefont {C.}~\bibnamefont
  {Schuster}}, \bibinfo {author} {\bibfnamefont {M.~J.}\ \bibnamefont
  {Shearn}}, \bibinfo {author} {\bibfnamefont {A.}~\bibnamefont {Shorter}},
  \bibinfo {author} {\bibfnamefont {V.}~\bibnamefont {Shvarts}}, \bibinfo
  {author} {\bibfnamefont {J.}~\bibnamefont {Skruzny}}, \bibinfo {author}
  {\bibfnamefont {V.}~\bibnamefont {Smelyanskiy}}, \bibinfo {author}
  {\bibfnamefont {W.~C.}\ \bibnamefont {Smith}}, \bibinfo {author}
  {\bibfnamefont {G.}~\bibnamefont {Sterling}}, \bibinfo {author}
  {\bibfnamefont {D.}~\bibnamefont {Strain}}, \bibinfo {author} {\bibfnamefont
  {M.}~\bibnamefont {Szalay}}, \bibinfo {author} {\bibfnamefont
  {A.}~\bibnamefont {Torres}}, \bibinfo {author} {\bibfnamefont
  {G.}~\bibnamefont {Vidal}}, \bibinfo {author} {\bibfnamefont
  {B.}~\bibnamefont {Villalonga}}, \bibinfo {author} {\bibfnamefont
  {C.}~\bibnamefont {Vollgraff~Heidweiller}}, \bibinfo {author} {\bibfnamefont
  {T.}~\bibnamefont {White}}, \bibinfo {author} {\bibfnamefont
  {C.}~\bibnamefont {Xing}}, \bibinfo {author} {\bibfnamefont {Z.~J.}\
  \bibnamefont {Yao}}, \bibinfo {author} {\bibfnamefont {P.}~\bibnamefont
  {Yeh}}, \bibinfo {author} {\bibfnamefont {J.}~\bibnamefont {Yoo}}, \bibinfo
  {author} {\bibfnamefont {G.}~\bibnamefont {Young}}, \bibinfo {author}
  {\bibfnamefont {A.}~\bibnamefont {Zalcman}}, \bibinfo {author} {\bibfnamefont
  {Y.}~\bibnamefont {Zhang}}, \bibinfo {author} {\bibfnamefont
  {N.}~\bibnamefont {Zhu}}, \ and\ \bibinfo {author} {\bibfnamefont {G.~Q.}\
  \bibnamefont {AI}},\ }\href {\doibase 10.1038/s41586-022-05434-1} {\bibfield
  {journal} {\bibinfo  {journal} {Nature}\ }\textbf {\bibinfo {volume} {614}},\
  \bibinfo {pages} {676} (\bibinfo {year} {2023})}\BibitemShut {NoStop}%
\bibitem [{\citenamefont {Bao}\ \emph {et~al.}(2022)\citenamefont {Bao},
  \citenamefont {Deng}, \citenamefont {Ding}, \citenamefont {Gao},
  \citenamefont {Gao}, \citenamefont {Huang}, \citenamefont {Jiang},
  \citenamefont {Ku}, \citenamefont {Li}, \citenamefont {Ma}, \citenamefont
  {Ni}, \citenamefont {Qin}, \citenamefont {Song}, \citenamefont {Sun},
  \citenamefont {Tang}, \citenamefont {Wang}, \citenamefont {Wu}, \citenamefont
  {Xia}, \citenamefont {Yu}, \citenamefont {Zhang}, \citenamefont {Zhang},
  \citenamefont {Zhang}, \citenamefont {Zhou}, \citenamefont {Zhu},
  \citenamefont {Shi}, \citenamefont {Chen}, \citenamefont {Zhao},\ and\
  \citenamefont {Deng}}]{PhysRevLett.129.010502}%
  \BibitemOpen
  \bibfield  {author} {\bibinfo {author} {\bibfnamefont {F.}~\bibnamefont
  {Bao}}, \bibinfo {author} {\bibfnamefont {H.}~\bibnamefont {Deng}}, \bibinfo
  {author} {\bibfnamefont {D.}~\bibnamefont {Ding}}, \bibinfo {author}
  {\bibfnamefont {R.}~\bibnamefont {Gao}}, \bibinfo {author} {\bibfnamefont
  {X.}~\bibnamefont {Gao}}, \bibinfo {author} {\bibfnamefont {C.}~\bibnamefont
  {Huang}}, \bibinfo {author} {\bibfnamefont {X.}~\bibnamefont {Jiang}},
  \bibinfo {author} {\bibfnamefont {H.-S.}\ \bibnamefont {Ku}}, \bibinfo
  {author} {\bibfnamefont {Z.}~\bibnamefont {Li}}, \bibinfo {author}
  {\bibfnamefont {X.}~\bibnamefont {Ma}}, \bibinfo {author} {\bibfnamefont
  {X.}~\bibnamefont {Ni}}, \bibinfo {author} {\bibfnamefont {J.}~\bibnamefont
  {Qin}}, \bibinfo {author} {\bibfnamefont {Z.}~\bibnamefont {Song}}, \bibinfo
  {author} {\bibfnamefont {H.}~\bibnamefont {Sun}}, \bibinfo {author}
  {\bibfnamefont {C.}~\bibnamefont {Tang}}, \bibinfo {author} {\bibfnamefont
  {T.}~\bibnamefont {Wang}}, \bibinfo {author} {\bibfnamefont {F.}~\bibnamefont
  {Wu}}, \bibinfo {author} {\bibfnamefont {T.}~\bibnamefont {Xia}}, \bibinfo
  {author} {\bibfnamefont {W.}~\bibnamefont {Yu}}, \bibinfo {author}
  {\bibfnamefont {F.}~\bibnamefont {Zhang}}, \bibinfo {author} {\bibfnamefont
  {G.}~\bibnamefont {Zhang}}, \bibinfo {author} {\bibfnamefont
  {X.}~\bibnamefont {Zhang}}, \bibinfo {author} {\bibfnamefont
  {J.}~\bibnamefont {Zhou}}, \bibinfo {author} {\bibfnamefont {X.}~\bibnamefont
  {Zhu}}, \bibinfo {author} {\bibfnamefont {Y.}~\bibnamefont {Shi}}, \bibinfo
  {author} {\bibfnamefont {J.}~\bibnamefont {Chen}}, \bibinfo {author}
  {\bibfnamefont {H.-H.}\ \bibnamefont {Zhao}}, \ and\ \bibinfo {author}
  {\bibfnamefont {C.}~\bibnamefont {Deng}},\ }\href {\doibase
  10.1103/PhysRevLett.129.010502} {\bibfield  {journal} {\bibinfo  {journal}
  {Phys. Rev. Lett.}\ }\textbf {\bibinfo {volume} {129}},\ \bibinfo {pages}
  {010502} (\bibinfo {year} {2022})}\BibitemShut {NoStop}%
\bibitem [{\citenamefont {Ballance}\ \emph {et~al.}(2016)\citenamefont
  {Ballance}, \citenamefont {Harty}, \citenamefont {Linke}, \citenamefont
  {Sepiol},\ and\ \citenamefont {Lucas}}]{PhysRevLett.117.060504}%
  \BibitemOpen
  \bibfield  {author} {\bibinfo {author} {\bibfnamefont {C.~J.}\ \bibnamefont
  {Ballance}}, \bibinfo {author} {\bibfnamefont {T.~P.}\ \bibnamefont {Harty}},
  \bibinfo {author} {\bibfnamefont {N.~M.}\ \bibnamefont {Linke}}, \bibinfo
  {author} {\bibfnamefont {M.~A.}\ \bibnamefont {Sepiol}}, \ and\ \bibinfo
  {author} {\bibfnamefont {D.~M.}\ \bibnamefont {Lucas}},\ }\href {\doibase
  10.1103/PhysRevLett.117.060504} {\bibfield  {journal} {\bibinfo  {journal}
  {Phys. Rev. Lett.}\ }\textbf {\bibinfo {volume} {117}},\ \bibinfo {pages}
  {060504} (\bibinfo {year} {2016})}\BibitemShut {NoStop}%
\bibitem [{\citenamefont {Gaebler}\ \emph {et~al.}(2016)\citenamefont
  {Gaebler}, \citenamefont {Tan}, \citenamefont {Lin}, \citenamefont {Wan},
  \citenamefont {Bowler}, \citenamefont {Keith}, \citenamefont {Glancy},
  \citenamefont {Coakley}, \citenamefont {Knill}, \citenamefont {Leibfried},\
  and\ \citenamefont {Wineland}}]{PhysRevLett.117.060505}%
  \BibitemOpen
  \bibfield  {author} {\bibinfo {author} {\bibfnamefont {J.~P.}\ \bibnamefont
  {Gaebler}}, \bibinfo {author} {\bibfnamefont {T.~R.}\ \bibnamefont {Tan}},
  \bibinfo {author} {\bibfnamefont {Y.}~\bibnamefont {Lin}}, \bibinfo {author}
  {\bibfnamefont {Y.}~\bibnamefont {Wan}}, \bibinfo {author} {\bibfnamefont
  {R.}~\bibnamefont {Bowler}}, \bibinfo {author} {\bibfnamefont {A.~C.}\
  \bibnamefont {Keith}}, \bibinfo {author} {\bibfnamefont {S.}~\bibnamefont
  {Glancy}}, \bibinfo {author} {\bibfnamefont {K.}~\bibnamefont {Coakley}},
  \bibinfo {author} {\bibfnamefont {E.}~\bibnamefont {Knill}}, \bibinfo
  {author} {\bibfnamefont {D.}~\bibnamefont {Leibfried}}, \ and\ \bibinfo
  {author} {\bibfnamefont {D.~J.}\ \bibnamefont {Wineland}},\ }\href {\doibase
  10.1103/PhysRevLett.117.060505} {\bibfield  {journal} {\bibinfo  {journal}
  {Phys. Rev. Lett.}\ }\textbf {\bibinfo {volume} {117}},\ \bibinfo {pages}
  {060505} (\bibinfo {year} {2016})}\BibitemShut {NoStop}%
\bibitem [{\citenamefont {Bluvstein}\ \emph {et~al.}(2024)\citenamefont
  {Bluvstein}, \citenamefont {Evered}, \citenamefont {Geim}, \citenamefont
  {Li}, \citenamefont {Zhou}, \citenamefont {Manovitz}, \citenamefont {Ebadi},
  \citenamefont {Cain}, \citenamefont {Kalinowski}, \citenamefont {Hangleiter},
  \citenamefont {Bonilla~Ataides}, \citenamefont {Maskara}, \citenamefont
  {Cong}, \citenamefont {Gao}, \citenamefont {Sales~Rodriguez}, \citenamefont
  {Karolyshyn}, \citenamefont {Semeghini}, \citenamefont {Gullans},
  \citenamefont {Greiner}, \citenamefont {Vuleti{\'{c}}},\ and\ \citenamefont
  {Lukin}}]{Bluvstein2024}%
  \BibitemOpen
  \bibfield  {author} {\bibinfo {author} {\bibfnamefont {D.}~\bibnamefont
  {Bluvstein}}, \bibinfo {author} {\bibfnamefont {S.~J.}\ \bibnamefont
  {Evered}}, \bibinfo {author} {\bibfnamefont {A.~A.}\ \bibnamefont {Geim}},
  \bibinfo {author} {\bibfnamefont {S.~H.}\ \bibnamefont {Li}}, \bibinfo
  {author} {\bibfnamefont {H.}~\bibnamefont {Zhou}}, \bibinfo {author}
  {\bibfnamefont {T.}~\bibnamefont {Manovitz}}, \bibinfo {author}
  {\bibfnamefont {S.}~\bibnamefont {Ebadi}}, \bibinfo {author} {\bibfnamefont
  {M.}~\bibnamefont {Cain}}, \bibinfo {author} {\bibfnamefont {M.}~\bibnamefont
  {Kalinowski}}, \bibinfo {author} {\bibfnamefont {D.}~\bibnamefont
  {Hangleiter}}, \bibinfo {author} {\bibfnamefont {J.~P.}\ \bibnamefont
  {Bonilla~Ataides}}, \bibinfo {author} {\bibfnamefont {N.}~\bibnamefont
  {Maskara}}, \bibinfo {author} {\bibfnamefont {I.}~\bibnamefont {Cong}},
  \bibinfo {author} {\bibfnamefont {X.}~\bibnamefont {Gao}}, \bibinfo {author}
  {\bibfnamefont {P.}~\bibnamefont {Sales~Rodriguez}}, \bibinfo {author}
  {\bibfnamefont {T.}~\bibnamefont {Karolyshyn}}, \bibinfo {author}
  {\bibfnamefont {G.}~\bibnamefont {Semeghini}}, \bibinfo {author}
  {\bibfnamefont {M.~J.}\ \bibnamefont {Gullans}}, \bibinfo {author}
  {\bibfnamefont {M.}~\bibnamefont {Greiner}}, \bibinfo {author} {\bibfnamefont
  {V.}~\bibnamefont {Vuleti{\'{c}}}}, \ and\ \bibinfo {author} {\bibfnamefont
  {M.~D.}\ \bibnamefont {Lukin}},\ }\href {\doibase 10.1038/s41586-023-06927-3}
  {\bibfield  {journal} {\bibinfo  {journal} {Nature}\ }\textbf {\bibinfo
  {volume} {626}},\ \bibinfo {pages} {58} (\bibinfo {year} {2024})}\BibitemShut
  {NoStop}%
\bibitem [{\citenamefont {Rong}\ \emph {et~al.}(2015)\citenamefont {Rong},
  \citenamefont {Geng}, \citenamefont {Shi}, \citenamefont {Liu}, \citenamefont
  {Xu}, \citenamefont {Ma}, \citenamefont {Kong}, \citenamefont {Jiang},
  \citenamefont {Wu},\ and\ \citenamefont {Du}}]{Rong2015}%
  \BibitemOpen
  \bibfield  {author} {\bibinfo {author} {\bibfnamefont {X.}~\bibnamefont
  {Rong}}, \bibinfo {author} {\bibfnamefont {J.}~\bibnamefont {Geng}}, \bibinfo
  {author} {\bibfnamefont {F.}~\bibnamefont {Shi}}, \bibinfo {author}
  {\bibfnamefont {Y.}~\bibnamefont {Liu}}, \bibinfo {author} {\bibfnamefont
  {K.}~\bibnamefont {Xu}}, \bibinfo {author} {\bibfnamefont {W.}~\bibnamefont
  {Ma}}, \bibinfo {author} {\bibfnamefont {F.}~\bibnamefont {Kong}}, \bibinfo
  {author} {\bibfnamefont {Z.}~\bibnamefont {Jiang}}, \bibinfo {author}
  {\bibfnamefont {Y.}~\bibnamefont {Wu}}, \ and\ \bibinfo {author}
  {\bibfnamefont {J.}~\bibnamefont {Du}},\ }\href {\doibase 10.1038/ncomms9748}
  {\bibfield  {journal} {\bibinfo  {journal} {Nature Communications}\ }\textbf
  {\bibinfo {volume} {6}},\ \bibinfo {pages} {8748} (\bibinfo {year}
  {2015})}\BibitemShut {NoStop}%
\bibitem [{\citenamefont {Noiri}\ \emph {et~al.}(2022)\citenamefont {Noiri},
  \citenamefont {Takeda}, \citenamefont {Nakajima}, \citenamefont {Kobayashi},
  \citenamefont {Sammak}, \citenamefont {Scappucci},\ and\ \citenamefont
  {Tarucha}}]{Noiri2022}%
  \BibitemOpen
  \bibfield  {author} {\bibinfo {author} {\bibfnamefont {A.}~\bibnamefont
  {Noiri}}, \bibinfo {author} {\bibfnamefont {K.}~\bibnamefont {Takeda}},
  \bibinfo {author} {\bibfnamefont {T.}~\bibnamefont {Nakajima}}, \bibinfo
  {author} {\bibfnamefont {T.}~\bibnamefont {Kobayashi}}, \bibinfo {author}
  {\bibfnamefont {A.}~\bibnamefont {Sammak}}, \bibinfo {author} {\bibfnamefont
  {G.}~\bibnamefont {Scappucci}}, \ and\ \bibinfo {author} {\bibfnamefont
  {S.}~\bibnamefont {Tarucha}},\ }\href {\doibase 10.1038/s41586-021-04182-y}
  {\bibfield  {journal} {\bibinfo  {journal} {Nature}\ }\textbf {\bibinfo
  {volume} {601}},\ \bibinfo {pages} {338} (\bibinfo {year}
  {2022})}\BibitemShut {NoStop}%
\bibitem [{\citenamefont {Wu}\ \emph {et~al.}(2021{\natexlab{a}})\citenamefont
  {Wu}, \citenamefont {Bao}, \citenamefont {Cao}, \citenamefont {Chen},
  \citenamefont {Chen}, \citenamefont {Chen}, \citenamefont {Chung},
  \citenamefont {Deng}, \citenamefont {Du}, \citenamefont {Fan}, \citenamefont
  {Gong}, \citenamefont {Guo}, \citenamefont {Guo}, \citenamefont {Guo},
  \citenamefont {Han}, \citenamefont {Hong}, \citenamefont {Huang},
  \citenamefont {Huo}, \citenamefont {Li}, \citenamefont {Li}, \citenamefont
  {Li}, \citenamefont {Li}, \citenamefont {Liang}, \citenamefont {Lin},
  \citenamefont {Lin}, \citenamefont {Qian}, \citenamefont {Qiao},
  \citenamefont {Rong}, \citenamefont {Su}, \citenamefont {Sun}, \citenamefont
  {Wang}, \citenamefont {Wang}, \citenamefont {Wu}, \citenamefont {Xu},
  \citenamefont {Yan}, \citenamefont {Yang}, \citenamefont {Yang},
  \citenamefont {Ye}, \citenamefont {Yin}, \citenamefont {Ying}, \citenamefont
  {Yu}, \citenamefont {Zha}, \citenamefont {Zhang}, \citenamefont {Zhang},
  \citenamefont {Zhang}, \citenamefont {Zhang}, \citenamefont {Zhao},
  \citenamefont {Zhao}, \citenamefont {Zhou}, \citenamefont {Zhu},
  \citenamefont {Lu}, \citenamefont {Peng}, \citenamefont {Zhu},\ and\
  \citenamefont {Pan}}]{PhysRevLett.127.180501}%
  \BibitemOpen
  \bibfield  {author} {\bibinfo {author} {\bibfnamefont {Y.}~\bibnamefont
  {Wu}}, \bibinfo {author} {\bibfnamefont {W.-S.}\ \bibnamefont {Bao}},
  \bibinfo {author} {\bibfnamefont {S.}~\bibnamefont {Cao}}, \bibinfo {author}
  {\bibfnamefont {F.}~\bibnamefont {Chen}}, \bibinfo {author} {\bibfnamefont
  {M.-C.}\ \bibnamefont {Chen}}, \bibinfo {author} {\bibfnamefont
  {X.}~\bibnamefont {Chen}}, \bibinfo {author} {\bibfnamefont {T.-H.}\
  \bibnamefont {Chung}}, \bibinfo {author} {\bibfnamefont {H.}~\bibnamefont
  {Deng}}, \bibinfo {author} {\bibfnamefont {Y.}~\bibnamefont {Du}}, \bibinfo
  {author} {\bibfnamefont {D.}~\bibnamefont {Fan}}, \bibinfo {author}
  {\bibfnamefont {M.}~\bibnamefont {Gong}}, \bibinfo {author} {\bibfnamefont
  {C.}~\bibnamefont {Guo}}, \bibinfo {author} {\bibfnamefont {C.}~\bibnamefont
  {Guo}}, \bibinfo {author} {\bibfnamefont {S.}~\bibnamefont {Guo}}, \bibinfo
  {author} {\bibfnamefont {L.}~\bibnamefont {Han}}, \bibinfo {author}
  {\bibfnamefont {L.}~\bibnamefont {Hong}}, \bibinfo {author} {\bibfnamefont
  {H.-L.}\ \bibnamefont {Huang}}, \bibinfo {author} {\bibfnamefont {Y.-H.}\
  \bibnamefont {Huo}}, \bibinfo {author} {\bibfnamefont {L.}~\bibnamefont
  {Li}}, \bibinfo {author} {\bibfnamefont {N.}~\bibnamefont {Li}}, \bibinfo
  {author} {\bibfnamefont {S.}~\bibnamefont {Li}}, \bibinfo {author}
  {\bibfnamefont {Y.}~\bibnamefont {Li}}, \bibinfo {author} {\bibfnamefont
  {F.}~\bibnamefont {Liang}}, \bibinfo {author} {\bibfnamefont
  {C.}~\bibnamefont {Lin}}, \bibinfo {author} {\bibfnamefont {J.}~\bibnamefont
  {Lin}}, \bibinfo {author} {\bibfnamefont {H.}~\bibnamefont {Qian}}, \bibinfo
  {author} {\bibfnamefont {D.}~\bibnamefont {Qiao}}, \bibinfo {author}
  {\bibfnamefont {H.}~\bibnamefont {Rong}}, \bibinfo {author} {\bibfnamefont
  {H.}~\bibnamefont {Su}}, \bibinfo {author} {\bibfnamefont {L.}~\bibnamefont
  {Sun}}, \bibinfo {author} {\bibfnamefont {L.}~\bibnamefont {Wang}}, \bibinfo
  {author} {\bibfnamefont {S.}~\bibnamefont {Wang}}, \bibinfo {author}
  {\bibfnamefont {D.}~\bibnamefont {Wu}}, \bibinfo {author} {\bibfnamefont
  {Y.}~\bibnamefont {Xu}}, \bibinfo {author} {\bibfnamefont {K.}~\bibnamefont
  {Yan}}, \bibinfo {author} {\bibfnamefont {W.}~\bibnamefont {Yang}}, \bibinfo
  {author} {\bibfnamefont {Y.}~\bibnamefont {Yang}}, \bibinfo {author}
  {\bibfnamefont {Y.}~\bibnamefont {Ye}}, \bibinfo {author} {\bibfnamefont
  {J.}~\bibnamefont {Yin}}, \bibinfo {author} {\bibfnamefont {C.}~\bibnamefont
  {Ying}}, \bibinfo {author} {\bibfnamefont {J.}~\bibnamefont {Yu}}, \bibinfo
  {author} {\bibfnamefont {C.}~\bibnamefont {Zha}}, \bibinfo {author}
  {\bibfnamefont {C.}~\bibnamefont {Zhang}}, \bibinfo {author} {\bibfnamefont
  {H.}~\bibnamefont {Zhang}}, \bibinfo {author} {\bibfnamefont
  {K.}~\bibnamefont {Zhang}}, \bibinfo {author} {\bibfnamefont
  {Y.}~\bibnamefont {Zhang}}, \bibinfo {author} {\bibfnamefont
  {H.}~\bibnamefont {Zhao}}, \bibinfo {author} {\bibfnamefont {Y.}~\bibnamefont
  {Zhao}}, \bibinfo {author} {\bibfnamefont {L.}~\bibnamefont {Zhou}}, \bibinfo
  {author} {\bibfnamefont {Q.}~\bibnamefont {Zhu}}, \bibinfo {author}
  {\bibfnamefont {C.-Y.}\ \bibnamefont {Lu}}, \bibinfo {author} {\bibfnamefont
  {C.-Z.}\ \bibnamefont {Peng}}, \bibinfo {author} {\bibfnamefont
  {X.}~\bibnamefont {Zhu}}, \ and\ \bibinfo {author} {\bibfnamefont {J.-W.}\
  \bibnamefont {Pan}},\ }\href {\doibase 10.1103/PhysRevLett.127.180501}
  {\bibfield  {journal} {\bibinfo  {journal} {Phys. Rev. Lett.}\ }\textbf
  {\bibinfo {volume} {127}},\ \bibinfo {pages} {180501} (\bibinfo {year}
  {2021}{\natexlab{a}})}\BibitemShut {NoStop}%
\bibitem [{\citenamefont {Moses}\ \emph {et~al.}(2023)\citenamefont {Moses},
  \citenamefont {Baldwin}, \citenamefont {Allman}, \citenamefont {Ancona},
  \citenamefont {Ascarrunz}, \citenamefont {Barnes}, \citenamefont
  {Bartolotta}, \citenamefont {Bjork}, \citenamefont {Blanchard}, \citenamefont
  {Bohn}, \citenamefont {Bohnet}, \citenamefont {Brown}, \citenamefont
  {Burdick}, \citenamefont {Burton}, \citenamefont {Campbell}, \citenamefont
  {Campora}, \citenamefont {Carron}, \citenamefont {Chambers}, \citenamefont
  {Chan}, \citenamefont {Chen}, \citenamefont {Chernoguzov}, \citenamefont
  {Chertkov}, \citenamefont {Colina}, \citenamefont {Curtis}, \citenamefont
  {Daniel}, \citenamefont {DeCross}, \citenamefont {Deen}, \citenamefont
  {Delaney}, \citenamefont {Dreiling}, \citenamefont {Ertsgaard}, \citenamefont
  {Esposito}, \citenamefont {Estey}, \citenamefont {Fabrikant}, \citenamefont
  {Figgatt}, \citenamefont {Foltz}, \citenamefont {Foss-Feig}, \citenamefont
  {Francois}, \citenamefont {Gaebler}, \citenamefont {Gatterman}, \citenamefont
  {Gilbreth}, \citenamefont {Giles}, \citenamefont {Glynn}, \citenamefont
  {Hall}, \citenamefont {Hankin}, \citenamefont {Hansen}, \citenamefont
  {Hayes}, \citenamefont {Higashi}, \citenamefont {Hoffman}, \citenamefont
  {Horning}, \citenamefont {Hout}, \citenamefont {Jacobs}, \citenamefont
  {Johansen}, \citenamefont {Jones}, \citenamefont {Karcz}, \citenamefont
  {Klein}, \citenamefont {Lauria}, \citenamefont {Lee}, \citenamefont {Liefer},
  \citenamefont {Lu}, \citenamefont {Lucchetti}, \citenamefont {Lytle},
  \citenamefont {Malm}, \citenamefont {Matheny}, \citenamefont {Mathewson},
  \citenamefont {Mayer}, \citenamefont {Miller}, \citenamefont {Mills},
  \citenamefont {Neyenhuis}, \citenamefont {Nugent}, \citenamefont {Olson},
  \citenamefont {Parks}, \citenamefont {Price}, \citenamefont {Price},
  \citenamefont {Pugh}, \citenamefont {Ransford}, \citenamefont {Reed},
  \citenamefont {Roman}, \citenamefont {Rowe}, \citenamefont {Ryan-Anderson},
  \citenamefont {Sanders}, \citenamefont {Sedlacek}, \citenamefont {Shevchuk},
  \citenamefont {Siegfried}, \citenamefont {Skripka}, \citenamefont {Spaun},
  \citenamefont {Sprenkle}, \citenamefont {Stutz}, \citenamefont {Swallows},
  \citenamefont {Tobey}, \citenamefont {Tran}, \citenamefont {Tran},
  \citenamefont {Vogt}, \citenamefont {Volin}, \citenamefont {Walker},
  \citenamefont {Zolot},\ and\ \citenamefont {Pino}}]{PhysRevX.13.041052}%
  \BibitemOpen
  \bibfield  {author} {\bibinfo {author} {\bibfnamefont {S.~A.}\ \bibnamefont
  {Moses}}, \bibinfo {author} {\bibfnamefont {C.~H.}\ \bibnamefont {Baldwin}},
  \bibinfo {author} {\bibfnamefont {M.~S.}\ \bibnamefont {Allman}}, \bibinfo
  {author} {\bibfnamefont {R.}~\bibnamefont {Ancona}}, \bibinfo {author}
  {\bibfnamefont {L.}~\bibnamefont {Ascarrunz}}, \bibinfo {author}
  {\bibfnamefont {C.}~\bibnamefont {Barnes}}, \bibinfo {author} {\bibfnamefont
  {J.}~\bibnamefont {Bartolotta}}, \bibinfo {author} {\bibfnamefont
  {B.}~\bibnamefont {Bjork}}, \bibinfo {author} {\bibfnamefont
  {P.}~\bibnamefont {Blanchard}}, \bibinfo {author} {\bibfnamefont
  {M.}~\bibnamefont {Bohn}}, \bibinfo {author} {\bibfnamefont {J.~G.}\
  \bibnamefont {Bohnet}}, \bibinfo {author} {\bibfnamefont {N.~C.}\
  \bibnamefont {Brown}}, \bibinfo {author} {\bibfnamefont {N.~Q.}\ \bibnamefont
  {Burdick}}, \bibinfo {author} {\bibfnamefont {W.~C.}\ \bibnamefont {Burton}},
  \bibinfo {author} {\bibfnamefont {S.~L.}\ \bibnamefont {Campbell}}, \bibinfo
  {author} {\bibfnamefont {J.~P.}\ \bibnamefont {Campora}}, \bibinfo {author}
  {\bibfnamefont {C.}~\bibnamefont {Carron}}, \bibinfo {author} {\bibfnamefont
  {J.}~\bibnamefont {Chambers}}, \bibinfo {author} {\bibfnamefont {J.~W.}\
  \bibnamefont {Chan}}, \bibinfo {author} {\bibfnamefont {Y.~H.}\ \bibnamefont
  {Chen}}, \bibinfo {author} {\bibfnamefont {A.}~\bibnamefont {Chernoguzov}},
  \bibinfo {author} {\bibfnamefont {E.}~\bibnamefont {Chertkov}}, \bibinfo
  {author} {\bibfnamefont {J.}~\bibnamefont {Colina}}, \bibinfo {author}
  {\bibfnamefont {J.~P.}\ \bibnamefont {Curtis}}, \bibinfo {author}
  {\bibfnamefont {R.}~\bibnamefont {Daniel}}, \bibinfo {author} {\bibfnamefont
  {M.}~\bibnamefont {DeCross}}, \bibinfo {author} {\bibfnamefont
  {D.}~\bibnamefont {Deen}}, \bibinfo {author} {\bibfnamefont {C.}~\bibnamefont
  {Delaney}}, \bibinfo {author} {\bibfnamefont {J.~M.}\ \bibnamefont
  {Dreiling}}, \bibinfo {author} {\bibfnamefont {C.~T.}\ \bibnamefont
  {Ertsgaard}}, \bibinfo {author} {\bibfnamefont {J.}~\bibnamefont {Esposito}},
  \bibinfo {author} {\bibfnamefont {B.}~\bibnamefont {Estey}}, \bibinfo
  {author} {\bibfnamefont {M.}~\bibnamefont {Fabrikant}}, \bibinfo {author}
  {\bibfnamefont {C.}~\bibnamefont {Figgatt}}, \bibinfo {author} {\bibfnamefont
  {C.}~\bibnamefont {Foltz}}, \bibinfo {author} {\bibfnamefont
  {M.}~\bibnamefont {Foss-Feig}}, \bibinfo {author} {\bibfnamefont
  {D.}~\bibnamefont {Francois}}, \bibinfo {author} {\bibfnamefont {J.~P.}\
  \bibnamefont {Gaebler}}, \bibinfo {author} {\bibfnamefont {T.~M.}\
  \bibnamefont {Gatterman}}, \bibinfo {author} {\bibfnamefont {C.~N.}\
  \bibnamefont {Gilbreth}}, \bibinfo {author} {\bibfnamefont {J.}~\bibnamefont
  {Giles}}, \bibinfo {author} {\bibfnamefont {E.}~\bibnamefont {Glynn}},
  \bibinfo {author} {\bibfnamefont {A.}~\bibnamefont {Hall}}, \bibinfo {author}
  {\bibfnamefont {A.~M.}\ \bibnamefont {Hankin}}, \bibinfo {author}
  {\bibfnamefont {A.}~\bibnamefont {Hansen}}, \bibinfo {author} {\bibfnamefont
  {D.}~\bibnamefont {Hayes}}, \bibinfo {author} {\bibfnamefont
  {B.}~\bibnamefont {Higashi}}, \bibinfo {author} {\bibfnamefont {I.~M.}\
  \bibnamefont {Hoffman}}, \bibinfo {author} {\bibfnamefont {B.}~\bibnamefont
  {Horning}}, \bibinfo {author} {\bibfnamefont {J.~J.}\ \bibnamefont {Hout}},
  \bibinfo {author} {\bibfnamefont {R.}~\bibnamefont {Jacobs}}, \bibinfo
  {author} {\bibfnamefont {J.}~\bibnamefont {Johansen}}, \bibinfo {author}
  {\bibfnamefont {L.}~\bibnamefont {Jones}}, \bibinfo {author} {\bibfnamefont
  {J.}~\bibnamefont {Karcz}}, \bibinfo {author} {\bibfnamefont
  {T.}~\bibnamefont {Klein}}, \bibinfo {author} {\bibfnamefont
  {P.}~\bibnamefont {Lauria}}, \bibinfo {author} {\bibfnamefont
  {P.}~\bibnamefont {Lee}}, \bibinfo {author} {\bibfnamefont {D.}~\bibnamefont
  {Liefer}}, \bibinfo {author} {\bibfnamefont {S.~T.}\ \bibnamefont {Lu}},
  \bibinfo {author} {\bibfnamefont {D.}~\bibnamefont {Lucchetti}}, \bibinfo
  {author} {\bibfnamefont {C.}~\bibnamefont {Lytle}}, \bibinfo {author}
  {\bibfnamefont {A.}~\bibnamefont {Malm}}, \bibinfo {author} {\bibfnamefont
  {M.}~\bibnamefont {Matheny}}, \bibinfo {author} {\bibfnamefont
  {B.}~\bibnamefont {Mathewson}}, \bibinfo {author} {\bibfnamefont
  {K.}~\bibnamefont {Mayer}}, \bibinfo {author} {\bibfnamefont {D.~B.}\
  \bibnamefont {Miller}}, \bibinfo {author} {\bibfnamefont {M.}~\bibnamefont
  {Mills}}, \bibinfo {author} {\bibfnamefont {B.}~\bibnamefont {Neyenhuis}},
  \bibinfo {author} {\bibfnamefont {L.}~\bibnamefont {Nugent}}, \bibinfo
  {author} {\bibfnamefont {S.}~\bibnamefont {Olson}}, \bibinfo {author}
  {\bibfnamefont {J.}~\bibnamefont {Parks}}, \bibinfo {author} {\bibfnamefont
  {G.~N.}\ \bibnamefont {Price}}, \bibinfo {author} {\bibfnamefont
  {Z.}~\bibnamefont {Price}}, \bibinfo {author} {\bibfnamefont
  {M.}~\bibnamefont {Pugh}}, \bibinfo {author} {\bibfnamefont {A.}~\bibnamefont
  {Ransford}}, \bibinfo {author} {\bibfnamefont {A.~P.}\ \bibnamefont {Reed}},
  \bibinfo {author} {\bibfnamefont {C.}~\bibnamefont {Roman}}, \bibinfo
  {author} {\bibfnamefont {M.}~\bibnamefont {Rowe}}, \bibinfo {author}
  {\bibfnamefont {C.}~\bibnamefont {Ryan-Anderson}}, \bibinfo {author}
  {\bibfnamefont {S.}~\bibnamefont {Sanders}}, \bibinfo {author} {\bibfnamefont
  {J.}~\bibnamefont {Sedlacek}}, \bibinfo {author} {\bibfnamefont
  {P.}~\bibnamefont {Shevchuk}}, \bibinfo {author} {\bibfnamefont
  {P.}~\bibnamefont {Siegfried}}, \bibinfo {author} {\bibfnamefont
  {T.}~\bibnamefont {Skripka}}, \bibinfo {author} {\bibfnamefont
  {B.}~\bibnamefont {Spaun}}, \bibinfo {author} {\bibfnamefont {R.~T.}\
  \bibnamefont {Sprenkle}}, \bibinfo {author} {\bibfnamefont {R.~P.}\
  \bibnamefont {Stutz}}, \bibinfo {author} {\bibfnamefont {M.}~\bibnamefont
  {Swallows}}, \bibinfo {author} {\bibfnamefont {R.~I.}\ \bibnamefont {Tobey}},
  \bibinfo {author} {\bibfnamefont {A.}~\bibnamefont {Tran}}, \bibinfo {author}
  {\bibfnamefont {T.}~\bibnamefont {Tran}}, \bibinfo {author} {\bibfnamefont
  {E.}~\bibnamefont {Vogt}}, \bibinfo {author} {\bibfnamefont {C.}~\bibnamefont
  {Volin}}, \bibinfo {author} {\bibfnamefont {J.}~\bibnamefont {Walker}},
  \bibinfo {author} {\bibfnamefont {A.~M.}\ \bibnamefont {Zolot}}, \ and\
  \bibinfo {author} {\bibfnamefont {J.~M.}\ \bibnamefont {Pino}},\ }\href
  {\doibase 10.1103/PhysRevX.13.041052} {\bibfield  {journal} {\bibinfo
  {journal} {Phys. Rev. X}\ }\textbf {\bibinfo {volume} {13}},\ \bibinfo
  {pages} {041052} (\bibinfo {year} {2023})}\BibitemShut {NoStop}%
\bibitem [{\citenamefont {Chen}\ \emph {et~al.}(2023)\citenamefont {Chen},
  \citenamefont {Nielsen}, \citenamefont {Ebert}, \citenamefont {Inlek},
  \citenamefont {Wright}, \citenamefont {Chaplin}, \citenamefont {Maksymov},
  \citenamefont {Páez}, \citenamefont {Poudel}, \citenamefont {Maunz},\ and\
  \citenamefont {Gamble}}]{chen2023benchmarking}%
  \BibitemOpen
  \bibfield  {author} {\bibinfo {author} {\bibfnamefont {J.-S.}\ \bibnamefont
  {Chen}}, \bibinfo {author} {\bibfnamefont {E.}~\bibnamefont {Nielsen}},
  \bibinfo {author} {\bibfnamefont {M.}~\bibnamefont {Ebert}}, \bibinfo
  {author} {\bibfnamefont {V.}~\bibnamefont {Inlek}}, \bibinfo {author}
  {\bibfnamefont {K.}~\bibnamefont {Wright}}, \bibinfo {author} {\bibfnamefont
  {V.}~\bibnamefont {Chaplin}}, \bibinfo {author} {\bibfnamefont
  {A.}~\bibnamefont {Maksymov}}, \bibinfo {author} {\bibfnamefont
  {E.}~\bibnamefont {Páez}}, \bibinfo {author} {\bibfnamefont
  {A.}~\bibnamefont {Poudel}}, \bibinfo {author} {\bibfnamefont
  {P.}~\bibnamefont {Maunz}}, \ and\ \bibinfo {author} {\bibfnamefont
  {J.}~\bibnamefont {Gamble}},\ }\href
  {https://doi.org/10.48550/arXiv.2308.05071} {\bibfield  {journal} {\bibinfo
  {journal} {arXiv:2308.05071}\ } (\bibinfo {year} {2023})}\BibitemShut
  {NoStop}%
\bibitem [{\citenamefont {Fowler}\ \emph {et~al.}(2012)\citenamefont {Fowler},
  \citenamefont {Mariantoni}, \citenamefont {Martinis},\ and\ \citenamefont
  {Cleland}}]{PhysRevA.86.032324}%
  \BibitemOpen
  \bibfield  {author} {\bibinfo {author} {\bibfnamefont {A.~G.}\ \bibnamefont
  {Fowler}}, \bibinfo {author} {\bibfnamefont {M.}~\bibnamefont {Mariantoni}},
  \bibinfo {author} {\bibfnamefont {J.~M.}\ \bibnamefont {Martinis}}, \ and\
  \bibinfo {author} {\bibfnamefont {A.~N.}\ \bibnamefont {Cleland}},\ }\href
  {\doibase 10.1103/PhysRevA.86.032324} {\bibfield  {journal} {\bibinfo
  {journal} {Phys. Rev. A}\ }\textbf {\bibinfo {volume} {86}},\ \bibinfo
  {pages} {032324} (\bibinfo {year} {2012})}\BibitemShut {NoStop}%
\bibitem [{\citenamefont {Bruzewicz}\ \emph {et~al.}(2019)\citenamefont
  {Bruzewicz}, \citenamefont {Chiaverini}, \citenamefont {McConnell},\ and\
  \citenamefont {Sage}}]{10.1063/1.5088164}%
  \BibitemOpen
  \bibfield  {author} {\bibinfo {author} {\bibfnamefont {C.~D.}\ \bibnamefont
  {Bruzewicz}}, \bibinfo {author} {\bibfnamefont {J.}~\bibnamefont
  {Chiaverini}}, \bibinfo {author} {\bibfnamefont {R.}~\bibnamefont
  {McConnell}}, \ and\ \bibinfo {author} {\bibfnamefont {J.~M.}\ \bibnamefont
  {Sage}},\ }\href {\doibase 10.1063/1.5088164} {\bibfield  {journal} {\bibinfo
   {journal} {Applied Physics Reviews}\ }\textbf {\bibinfo {volume} {6}},\
  \bibinfo {pages} {021314} (\bibinfo {year} {2019})}\BibitemShut {NoStop}%
\bibitem [{\citenamefont {Harty}\ \emph {et~al.}(2014)\citenamefont {Harty},
  \citenamefont {Allcock}, \citenamefont {Ballance}, \citenamefont {Guidoni},
  \citenamefont {Janacek}, \citenamefont {Linke}, \citenamefont {Stacey},\ and\
  \citenamefont {Lucas}}]{PhysRevLett.113.220501}%
  \BibitemOpen
  \bibfield  {author} {\bibinfo {author} {\bibfnamefont {T.~P.}\ \bibnamefont
  {Harty}}, \bibinfo {author} {\bibfnamefont {D.~T.~C.}\ \bibnamefont
  {Allcock}}, \bibinfo {author} {\bibfnamefont {C.~J.}\ \bibnamefont
  {Ballance}}, \bibinfo {author} {\bibfnamefont {L.}~\bibnamefont {Guidoni}},
  \bibinfo {author} {\bibfnamefont {H.~A.}\ \bibnamefont {Janacek}}, \bibinfo
  {author} {\bibfnamefont {N.~M.}\ \bibnamefont {Linke}}, \bibinfo {author}
  {\bibfnamefont {D.~N.}\ \bibnamefont {Stacey}}, \ and\ \bibinfo {author}
  {\bibfnamefont {D.~M.}\ \bibnamefont {Lucas}},\ }\href {\doibase
  10.1103/PhysRevLett.113.220501} {\bibfield  {journal} {\bibinfo  {journal}
  {Phys. Rev. Lett.}\ }\textbf {\bibinfo {volume} {113}},\ \bibinfo {pages}
  {220501} (\bibinfo {year} {2014})}\BibitemShut {NoStop}%
\bibitem [{\citenamefont {Clark}\ \emph {et~al.}(2021)\citenamefont {Clark},
  \citenamefont {Tinkey}, \citenamefont {Sawyer}, \citenamefont {Meier},
  \citenamefont {Burkhardt}, \citenamefont {Seck}, \citenamefont {Shappert},
  \citenamefont {Guise}, \citenamefont {Volin}, \citenamefont {Fallek},
  \citenamefont {Hayden}, \citenamefont {Rellergert},\ and\ \citenamefont
  {Brown}}]{PhysRevLett.127.130505}%
  \BibitemOpen
  \bibfield  {author} {\bibinfo {author} {\bibfnamefont {C.~R.}\ \bibnamefont
  {Clark}}, \bibinfo {author} {\bibfnamefont {H.~N.}\ \bibnamefont {Tinkey}},
  \bibinfo {author} {\bibfnamefont {B.~C.}\ \bibnamefont {Sawyer}}, \bibinfo
  {author} {\bibfnamefont {A.~M.}\ \bibnamefont {Meier}}, \bibinfo {author}
  {\bibfnamefont {K.~A.}\ \bibnamefont {Burkhardt}}, \bibinfo {author}
  {\bibfnamefont {C.~M.}\ \bibnamefont {Seck}}, \bibinfo {author}
  {\bibfnamefont {C.~M.}\ \bibnamefont {Shappert}}, \bibinfo {author}
  {\bibfnamefont {N.~D.}\ \bibnamefont {Guise}}, \bibinfo {author}
  {\bibfnamefont {C.~E.}\ \bibnamefont {Volin}}, \bibinfo {author}
  {\bibfnamefont {S.~D.}\ \bibnamefont {Fallek}}, \bibinfo {author}
  {\bibfnamefont {H.~T.}\ \bibnamefont {Hayden}}, \bibinfo {author}
  {\bibfnamefont {W.~G.}\ \bibnamefont {Rellergert}}, \ and\ \bibinfo {author}
  {\bibfnamefont {K.~R.}\ \bibnamefont {Brown}},\ }\href {\doibase
  10.1103/PhysRevLett.127.130505} {\bibfield  {journal} {\bibinfo  {journal}
  {Phys. Rev. Lett.}\ }\textbf {\bibinfo {volume} {127}},\ \bibinfo {pages}
  {130505} (\bibinfo {year} {2021})}\BibitemShut {NoStop}%
\bibitem [{\citenamefont {An}\ \emph {et~al.}(2022)\citenamefont {An},
  \citenamefont {Ransford}, \citenamefont {Schaffer}, \citenamefont {Sletten},
  \citenamefont {Gaebler}, \citenamefont {Hostetter},\ and\ \citenamefont
  {Vittorini}}]{PhysRevLett.129.130501}%
  \BibitemOpen
  \bibfield  {author} {\bibinfo {author} {\bibfnamefont {F.~A.}\ \bibnamefont
  {An}}, \bibinfo {author} {\bibfnamefont {A.}~\bibnamefont {Ransford}},
  \bibinfo {author} {\bibfnamefont {A.}~\bibnamefont {Schaffer}}, \bibinfo
  {author} {\bibfnamefont {L.~R.}\ \bibnamefont {Sletten}}, \bibinfo {author}
  {\bibfnamefont {J.}~\bibnamefont {Gaebler}}, \bibinfo {author} {\bibfnamefont
  {J.}~\bibnamefont {Hostetter}}, \ and\ \bibinfo {author} {\bibfnamefont
  {G.}~\bibnamefont {Vittorini}},\ }\href {\doibase
  10.1103/PhysRevLett.129.130501} {\bibfield  {journal} {\bibinfo  {journal}
  {Phys. Rev. Lett.}\ }\textbf {\bibinfo {volume} {129}},\ \bibinfo {pages}
  {130501} (\bibinfo {year} {2022})}\BibitemShut {NoStop}%
\bibitem [{\citenamefont {Wineland}\ \emph {et~al.}(1998)\citenamefont
  {Wineland}, \citenamefont {Monroe}, \citenamefont {Itano}, \citenamefont
  {Leibfried}, \citenamefont {King},\ and\ \citenamefont
  {Meekhof}}]{wineland1998experimental}%
  \BibitemOpen
  \bibfield  {author} {\bibinfo {author} {\bibfnamefont {D.~J.}\ \bibnamefont
  {Wineland}}, \bibinfo {author} {\bibfnamefont {C.}~\bibnamefont {Monroe}},
  \bibinfo {author} {\bibfnamefont {W.~M.}\ \bibnamefont {Itano}}, \bibinfo
  {author} {\bibfnamefont {D.}~\bibnamefont {Leibfried}}, \bibinfo {author}
  {\bibfnamefont {B.~E.}\ \bibnamefont {King}}, \ and\ \bibinfo {author}
  {\bibfnamefont {D.~M.}\ \bibnamefont {Meekhof}},\ }\href
  {https://doi.org/10.6028%2Fjres.103.019} {\bibfield  {journal} {\bibinfo
  {journal} {Journal of research of the National Institute of Standards and
  Technology}\ }\textbf {\bibinfo {volume} {103}},\ \bibinfo {pages} {259}
  (\bibinfo {year} {1998})}\BibitemShut {NoStop}%
\bibitem [{\citenamefont {Hughes}\ \emph {et~al.}(1996)\citenamefont {Hughes},
  \citenamefont {James}, \citenamefont {Knill}, \citenamefont {Laflamme},\ and\
  \citenamefont {Petschek}}]{PhysRevLett.77.3240}%
  \BibitemOpen
  \bibfield  {author} {\bibinfo {author} {\bibfnamefont {R.~J.}\ \bibnamefont
  {Hughes}}, \bibinfo {author} {\bibfnamefont {D.~F.~V.}\ \bibnamefont
  {James}}, \bibinfo {author} {\bibfnamefont {E.~H.}\ \bibnamefont {Knill}},
  \bibinfo {author} {\bibfnamefont {R.}~\bibnamefont {Laflamme}}, \ and\
  \bibinfo {author} {\bibfnamefont {A.~G.}\ \bibnamefont {Petschek}},\ }\href
  {\doibase 10.1103/PhysRevLett.77.3240} {\bibfield  {journal} {\bibinfo
  {journal} {Phys. Rev. Lett.}\ }\textbf {\bibinfo {volume} {77}},\ \bibinfo
  {pages} {3240} (\bibinfo {year} {1996})}\BibitemShut {NoStop}%
\bibitem [{\citenamefont {Clark}(2001)}]{clark2001proceedings}%
  \BibitemOpen
  \bibfield  {author} {\bibinfo {author} {\bibfnamefont {R.}~\bibnamefont
  {Clark}},\ }\href@noop {} {\emph {\bibinfo {title} {Proceedings of the 1st
  International Conference on Experimental Implementation of Quantum
  Computation: Sydney, Australia, 16-19 January 2001}}}\ (\bibinfo  {publisher}
  {Rinton Press},\ \bibinfo {year} {2001})\BibitemShut {NoStop}%
\bibitem [{\citenamefont {Kielpinski}\ \emph {et~al.}(2002)\citenamefont
  {Kielpinski}, \citenamefont {Monroe},\ and\ \citenamefont
  {Wineland}}]{Kielpinski2002}%
  \BibitemOpen
  \bibfield  {author} {\bibinfo {author} {\bibfnamefont {D.}~\bibnamefont
  {Kielpinski}}, \bibinfo {author} {\bibfnamefont {C.}~\bibnamefont {Monroe}},
  \ and\ \bibinfo {author} {\bibfnamefont {D.~J.}\ \bibnamefont {Wineland}},\
  }\href {\doibase 10.1038/nature00784} {\bibfield  {journal} {\bibinfo
  {journal} {Nature}\ }\textbf {\bibinfo {volume} {417}},\ \bibinfo {pages}
  {709} (\bibinfo {year} {2002})}\BibitemShut {NoStop}%
\bibitem [{\citenamefont {Duan}\ \emph {et~al.}(2004)\citenamefont {Duan},
  \citenamefont {Blinov}, \citenamefont {Moehring},\ and\ \citenamefont
  {Monroe}}]{10.5555/2011617.2011618}%
  \BibitemOpen
  \bibfield  {author} {\bibinfo {author} {\bibfnamefont {L.-M.}\ \bibnamefont
  {Duan}}, \bibinfo {author} {\bibfnamefont {B.~B.}\ \bibnamefont {Blinov}},
  \bibinfo {author} {\bibfnamefont {D.~L.}\ \bibnamefont {Moehring}}, \ and\
  \bibinfo {author} {\bibfnamefont {C.}~\bibnamefont {Monroe}},\ }\href@noop {}
  {\bibfield  {journal} {\bibinfo  {journal} {Quantum Info. Comput.}\ }\textbf
  {\bibinfo {volume} {4}},\ \bibinfo {pages} {165} (\bibinfo {year}
  {2004})}\BibitemShut {NoStop}%
\bibitem [{\citenamefont {Duan}\ and\ \citenamefont
  {Monroe}(2010)}]{RevModPhys.82.1209}%
  \BibitemOpen
  \bibfield  {author} {\bibinfo {author} {\bibfnamefont {L.-M.}\ \bibnamefont
  {Duan}}\ and\ \bibinfo {author} {\bibfnamefont {C.}~\bibnamefont {Monroe}},\
  }\href {\doibase 10.1103/RevModPhys.82.1209} {\bibfield  {journal} {\bibinfo
  {journal} {Rev. Mod. Phys.}\ }\textbf {\bibinfo {volume} {82}},\ \bibinfo
  {pages} {1209} (\bibinfo {year} {2010})}\BibitemShut {NoStop}%
\bibitem [{\citenamefont {Monroe}\ \emph {et~al.}(2014)\citenamefont {Monroe},
  \citenamefont {Raussendorf}, \citenamefont {Ruthven}, \citenamefont {Brown},
  \citenamefont {Maunz}, \citenamefont {Duan},\ and\ \citenamefont
  {Kim}}]{PhysRevA.89.022317}%
  \BibitemOpen
  \bibfield  {author} {\bibinfo {author} {\bibfnamefont {C.}~\bibnamefont
  {Monroe}}, \bibinfo {author} {\bibfnamefont {R.}~\bibnamefont {Raussendorf}},
  \bibinfo {author} {\bibfnamefont {A.}~\bibnamefont {Ruthven}}, \bibinfo
  {author} {\bibfnamefont {K.~R.}\ \bibnamefont {Brown}}, \bibinfo {author}
  {\bibfnamefont {P.}~\bibnamefont {Maunz}}, \bibinfo {author} {\bibfnamefont
  {L.-M.}\ \bibnamefont {Duan}}, \ and\ \bibinfo {author} {\bibfnamefont
  {J.}~\bibnamefont {Kim}},\ }\href {\doibase 10.1103/PhysRevA.89.022317}
  {\bibfield  {journal} {\bibinfo  {journal} {Phys. Rev. A}\ }\textbf {\bibinfo
  {volume} {89}},\ \bibinfo {pages} {022317} (\bibinfo {year}
  {2014})}\BibitemShut {NoStop}%
\bibitem [{\citenamefont {Stephenson}\ \emph {et~al.}(2020)\citenamefont
  {Stephenson}, \citenamefont {Nadlinger}, \citenamefont {Nichol},
  \citenamefont {An}, \citenamefont {Drmota}, \citenamefont {Ballance},
  \citenamefont {Thirumalai}, \citenamefont {Goodwin}, \citenamefont {Lucas},\
  and\ \citenamefont {Ballance}}]{PhysRevLett.124.110501}%
  \BibitemOpen
  \bibfield  {author} {\bibinfo {author} {\bibfnamefont {L.~J.}\ \bibnamefont
  {Stephenson}}, \bibinfo {author} {\bibfnamefont {D.~P.}\ \bibnamefont
  {Nadlinger}}, \bibinfo {author} {\bibfnamefont {B.~C.}\ \bibnamefont
  {Nichol}}, \bibinfo {author} {\bibfnamefont {S.}~\bibnamefont {An}}, \bibinfo
  {author} {\bibfnamefont {P.}~\bibnamefont {Drmota}}, \bibinfo {author}
  {\bibfnamefont {T.~G.}\ \bibnamefont {Ballance}}, \bibinfo {author}
  {\bibfnamefont {K.}~\bibnamefont {Thirumalai}}, \bibinfo {author}
  {\bibfnamefont {J.~F.}\ \bibnamefont {Goodwin}}, \bibinfo {author}
  {\bibfnamefont {D.~M.}\ \bibnamefont {Lucas}}, \ and\ \bibinfo {author}
  {\bibfnamefont {C.~J.}\ \bibnamefont {Ballance}},\ }\href {\doibase
  10.1103/PhysRevLett.124.110501} {\bibfield  {journal} {\bibinfo  {journal}
  {Phys. Rev. Lett.}\ }\textbf {\bibinfo {volume} {124}},\ \bibinfo {pages}
  {110501} (\bibinfo {year} {2020})}\BibitemShut {NoStop}%
\bibitem [{\citenamefont {Szymanski}\ \emph {et~al.}(2012)\citenamefont
  {Szymanski}, \citenamefont {Dubessy}, \citenamefont {Dubost}, \citenamefont
  {Guibal}, \citenamefont {Likforman},\ and\ \citenamefont
  {Guidoni}}]{Szymanski2012crystal}%
  \BibitemOpen
  \bibfield  {author} {\bibinfo {author} {\bibfnamefont {B.}~\bibnamefont
  {Szymanski}}, \bibinfo {author} {\bibfnamefont {R.}~\bibnamefont {Dubessy}},
  \bibinfo {author} {\bibfnamefont {B.}~\bibnamefont {Dubost}}, \bibinfo
  {author} {\bibfnamefont {S.}~\bibnamefont {Guibal}}, \bibinfo {author}
  {\bibfnamefont {J.-P.}\ \bibnamefont {Likforman}}, \ and\ \bibinfo {author}
  {\bibfnamefont {L.}~\bibnamefont {Guidoni}},\ }\href {\doibase
  10.1063/1.4705153} {\bibfield  {journal} {\bibinfo  {journal} {Applied
  Physics Letters}\ }\textbf {\bibinfo {volume} {100}},\ \bibinfo {pages}
  {171110} (\bibinfo {year} {2012})}\BibitemShut {NoStop}%
\bibitem [{\citenamefont {Wang}\ \emph
  {et~al.}(2020{\natexlab{a}})\citenamefont {Wang}, \citenamefont {Qiao},
  \citenamefont {Cai}, \citenamefont {Zhang}, \citenamefont {Jin},
  \citenamefont {Wang}, \citenamefont {Chen}, \citenamefont {Luan},
  \citenamefont {Du}, \citenamefont {Wang}, \citenamefont {Song}, \citenamefont
  {Yum},\ and\ \citenamefont {Kim}}]{https://doi.org/10.1002/qute.202000068}%
  \BibitemOpen
  \bibfield  {author} {\bibinfo {author} {\bibfnamefont {Y.}~\bibnamefont
  {Wang}}, \bibinfo {author} {\bibfnamefont {M.}~\bibnamefont {Qiao}}, \bibinfo
  {author} {\bibfnamefont {Z.}~\bibnamefont {Cai}}, \bibinfo {author}
  {\bibfnamefont {K.}~\bibnamefont {Zhang}}, \bibinfo {author} {\bibfnamefont
  {N.}~\bibnamefont {Jin}}, \bibinfo {author} {\bibfnamefont {P.}~\bibnamefont
  {Wang}}, \bibinfo {author} {\bibfnamefont {W.}~\bibnamefont {Chen}}, \bibinfo
  {author} {\bibfnamefont {C.}~\bibnamefont {Luan}}, \bibinfo {author}
  {\bibfnamefont {B.}~\bibnamefont {Du}}, \bibinfo {author} {\bibfnamefont
  {H.}~\bibnamefont {Wang}}, \bibinfo {author} {\bibfnamefont {Y.}~\bibnamefont
  {Song}}, \bibinfo {author} {\bibfnamefont {D.}~\bibnamefont {Yum}}, \ and\
  \bibinfo {author} {\bibfnamefont {K.}~\bibnamefont {Kim}},\ }\href {\doibase
  https://doi.org/10.1002/qute.202000068} {\bibfield  {journal} {\bibinfo
  {journal} {Advanced Quantum Technologies}\ }\textbf {\bibinfo {volume} {3}},\
  \bibinfo {pages} {2000068} (\bibinfo {year}
  {2020}{\natexlab{a}})}\BibitemShut {NoStop}%
\bibitem [{\citenamefont {Xie}\ \emph {et~al.}(2021)\citenamefont {Xie},
  \citenamefont {Cui}, \citenamefont {D’Onofrio}, \citenamefont {Rasmusson},
  \citenamefont {Howell},\ and\ \citenamefont {Richerme}}]{Xie_2021}%
  \BibitemOpen
  \bibfield  {author} {\bibinfo {author} {\bibfnamefont {Y.}~\bibnamefont
  {Xie}}, \bibinfo {author} {\bibfnamefont {J.}~\bibnamefont {Cui}}, \bibinfo
  {author} {\bibfnamefont {M.}~\bibnamefont {D’Onofrio}}, \bibinfo {author}
  {\bibfnamefont {A.~J.}\ \bibnamefont {Rasmusson}}, \bibinfo {author}
  {\bibfnamefont {S.~W.}\ \bibnamefont {Howell}}, \ and\ \bibinfo {author}
  {\bibfnamefont {P.}~\bibnamefont {Richerme}},\ }\href {\doibase
  10.1088/2058-9565/ac1e38} {\bibfield  {journal} {\bibinfo  {journal} {Quantum
  Science and Technology}\ }\textbf {\bibinfo {volume} {6}},\ \bibinfo {pages}
  {044009} (\bibinfo {year} {2021})}\BibitemShut {NoStop}%
\bibitem [{\citenamefont {Kato}\ \emph {et~al.}(2022)\citenamefont {Kato},
  \citenamefont {Goel}, \citenamefont {Lee}, \citenamefont {Ye}, \citenamefont
  {Karki}, \citenamefont {Liu}, \citenamefont {Nomerotski},\ and\ \citenamefont
  {Blinov}}]{PhysRevA.105.023101}%
  \BibitemOpen
  \bibfield  {author} {\bibinfo {author} {\bibfnamefont {A.}~\bibnamefont
  {Kato}}, \bibinfo {author} {\bibfnamefont {A.}~\bibnamefont {Goel}}, \bibinfo
  {author} {\bibfnamefont {R.}~\bibnamefont {Lee}}, \bibinfo {author}
  {\bibfnamefont {Z.}~\bibnamefont {Ye}}, \bibinfo {author} {\bibfnamefont
  {S.}~\bibnamefont {Karki}}, \bibinfo {author} {\bibfnamefont {J.~J.}\
  \bibnamefont {Liu}}, \bibinfo {author} {\bibfnamefont {A.}~\bibnamefont
  {Nomerotski}}, \ and\ \bibinfo {author} {\bibfnamefont {B.~B.}\ \bibnamefont
  {Blinov}},\ }\href {\doibase 10.1103/PhysRevA.105.023101} {\bibfield
  {journal} {\bibinfo  {journal} {Phys. Rev. A}\ }\textbf {\bibinfo {volume}
  {105}},\ \bibinfo {pages} {023101} (\bibinfo {year} {2022})}\BibitemShut
  {NoStop}%
\bibitem [{\citenamefont {Kiesenhofer}\ \emph {et~al.}(2023)\citenamefont
  {Kiesenhofer}, \citenamefont {Hainzer}, \citenamefont {Zhdanov},
  \citenamefont {Holz}, \citenamefont {Bock}, \citenamefont {Ollikainen},\ and\
  \citenamefont {Roos}}]{PRXQuantum.4.020317}%
  \BibitemOpen
  \bibfield  {author} {\bibinfo {author} {\bibfnamefont {D.}~\bibnamefont
  {Kiesenhofer}}, \bibinfo {author} {\bibfnamefont {H.}~\bibnamefont
  {Hainzer}}, \bibinfo {author} {\bibfnamefont {A.}~\bibnamefont {Zhdanov}},
  \bibinfo {author} {\bibfnamefont {P.~C.}\ \bibnamefont {Holz}}, \bibinfo
  {author} {\bibfnamefont {M.}~\bibnamefont {Bock}}, \bibinfo {author}
  {\bibfnamefont {T.}~\bibnamefont {Ollikainen}}, \ and\ \bibinfo {author}
  {\bibfnamefont {C.~F.}\ \bibnamefont {Roos}},\ }\href {\doibase
  10.1103/PRXQuantum.4.020317} {\bibfield  {journal} {\bibinfo  {journal} {PRX
  Quantum}\ }\textbf {\bibinfo {volume} {4}},\ \bibinfo {pages} {020317}
  (\bibinfo {year} {2023})}\BibitemShut {NoStop}%
\bibitem [{\citenamefont {Qiao}\ \emph {et~al.}(2024)\citenamefont {Qiao},
  \citenamefont {Cai}, \citenamefont {Wang}, \citenamefont {Du}, \citenamefont
  {Jin}, \citenamefont {Chen}, \citenamefont {Wang}, \citenamefont {Luan},
  \citenamefont {Gao}, \citenamefont {Sun}, \citenamefont {Tian}, \citenamefont
  {Zhang},\ and\ \citenamefont {Kim}}]{Qiao2024}%
  \BibitemOpen
  \bibfield  {author} {\bibinfo {author} {\bibfnamefont {M.}~\bibnamefont
  {Qiao}}, \bibinfo {author} {\bibfnamefont {Z.}~\bibnamefont {Cai}}, \bibinfo
  {author} {\bibfnamefont {Y.}~\bibnamefont {Wang}}, \bibinfo {author}
  {\bibfnamefont {B.}~\bibnamefont {Du}}, \bibinfo {author} {\bibfnamefont
  {N.}~\bibnamefont {Jin}}, \bibinfo {author} {\bibfnamefont {W.}~\bibnamefont
  {Chen}}, \bibinfo {author} {\bibfnamefont {P.}~\bibnamefont {Wang}}, \bibinfo
  {author} {\bibfnamefont {C.}~\bibnamefont {Luan}}, \bibinfo {author}
  {\bibfnamefont {E.}~\bibnamefont {Gao}}, \bibinfo {author} {\bibfnamefont
  {X.}~\bibnamefont {Sun}}, \bibinfo {author} {\bibfnamefont {H.}~\bibnamefont
  {Tian}}, \bibinfo {author} {\bibfnamefont {J.}~\bibnamefont {Zhang}}, \ and\
  \bibinfo {author} {\bibfnamefont {K.}~\bibnamefont {Kim}},\ }\href {\doibase
  10.1038/s41567-023-02378-9} {\bibfield  {journal} {\bibinfo  {journal}
  {Nature Physics}\ }\textbf {\bibinfo {volume} {20}},\ \bibinfo {pages} {623}
  (\bibinfo {year} {2024})}\BibitemShut {NoStop}%
\bibitem [{\citenamefont {Guo}\ \emph {et~al.}(2024)\citenamefont {Guo},
  \citenamefont {Wu}, \citenamefont {Ye}, \citenamefont {Zhang}, \citenamefont
  {Lian}, \citenamefont {Yao}, \citenamefont {Wang}, \citenamefont {Yan},
  \citenamefont {Yi}, \citenamefont {Xu}, \citenamefont {Li}, \citenamefont
  {Hou}, \citenamefont {Xu}, \citenamefont {Guo}, \citenamefont {Zhang},
  \citenamefont {Qi}, \citenamefont {Zhou}, \citenamefont {He},\ and\
  \citenamefont {Duan}}]{guo2023siteresolved}%
  \BibitemOpen
  \bibfield  {author} {\bibinfo {author} {\bibfnamefont {S.-A.}\ \bibnamefont
  {Guo}}, \bibinfo {author} {\bibfnamefont {Y.-K.}\ \bibnamefont {Wu}},
  \bibinfo {author} {\bibfnamefont {J.}~\bibnamefont {Ye}}, \bibinfo {author}
  {\bibfnamefont {L.}~\bibnamefont {Zhang}}, \bibinfo {author} {\bibfnamefont
  {W.-Q.}\ \bibnamefont {Lian}}, \bibinfo {author} {\bibfnamefont
  {R.}~\bibnamefont {Yao}}, \bibinfo {author} {\bibfnamefont {Y.}~\bibnamefont
  {Wang}}, \bibinfo {author} {\bibfnamefont {R.-Y.}\ \bibnamefont {Yan}},
  \bibinfo {author} {\bibfnamefont {Y.-J.}\ \bibnamefont {Yi}}, \bibinfo
  {author} {\bibfnamefont {Y.-L.}\ \bibnamefont {Xu}}, \bibinfo {author}
  {\bibfnamefont {B.-W.}\ \bibnamefont {Li}}, \bibinfo {author} {\bibfnamefont
  {Y.-H.}\ \bibnamefont {Hou}}, \bibinfo {author} {\bibfnamefont {Y.-Z.}\
  \bibnamefont {Xu}}, \bibinfo {author} {\bibfnamefont {W.-X.}\ \bibnamefont
  {Guo}}, \bibinfo {author} {\bibfnamefont {C.}~\bibnamefont {Zhang}}, \bibinfo
  {author} {\bibfnamefont {B.-X.}\ \bibnamefont {Qi}}, \bibinfo {author}
  {\bibfnamefont {Z.-C.}\ \bibnamefont {Zhou}}, \bibinfo {author}
  {\bibfnamefont {L.}~\bibnamefont {He}}, \ and\ \bibinfo {author}
  {\bibfnamefont {L.-M.}\ \bibnamefont {Duan}},\ }\href {\doibase
  10.1038/s41586-024-07459-0} {\bibfield  {journal} {\bibinfo  {journal}
  {Nature}\ }\textbf {\bibinfo {volume} {630}},\ \bibinfo {pages} {613}
  (\bibinfo {year} {2024})}\BibitemShut {NoStop}%
\bibitem [{\citenamefont {Shen}\ and\ \citenamefont
  {Duan}(2014)}]{PhysRevA.90.022332}%
  \BibitemOpen
  \bibfield  {author} {\bibinfo {author} {\bibfnamefont {C.}~\bibnamefont
  {Shen}}\ and\ \bibinfo {author} {\bibfnamefont {L.-M.}\ \bibnamefont
  {Duan}},\ }\href {\doibase 10.1103/PhysRevA.90.022332} {\bibfield  {journal}
  {\bibinfo  {journal} {Phys. Rev. A}\ }\textbf {\bibinfo {volume} {90}},\
  \bibinfo {pages} {022332} (\bibinfo {year} {2014})}\BibitemShut {NoStop}%
\bibitem [{\citenamefont {Wang}\ \emph {et~al.}(2015)\citenamefont {Wang},
  \citenamefont {Shen},\ and\ \citenamefont {Duan}}]{wang2015quantum}%
  \BibitemOpen
  \bibfield  {author} {\bibinfo {author} {\bibfnamefont {S.-T.}\ \bibnamefont
  {Wang}}, \bibinfo {author} {\bibfnamefont {C.}~\bibnamefont {Shen}}, \ and\
  \bibinfo {author} {\bibfnamefont {L.-M.}\ \bibnamefont {Duan}},\ }\href
  {https://doi.org/10.1038/srep08555} {\bibfield  {journal} {\bibinfo
  {journal} {Scientific reports}\ }\textbf {\bibinfo {volume} {5}},\ \bibinfo
  {pages} {8555} (\bibinfo {year} {2015})}\BibitemShut {NoStop}%
\bibitem [{\citenamefont {Bermudez}\ \emph {et~al.}(2017)\citenamefont
  {Bermudez}, \citenamefont {Schindler}, \citenamefont {Monz}, \citenamefont
  {Blatt},\ and\ \citenamefont {Müller}}]{Bermudez_2017}%
  \BibitemOpen
  \bibfield  {author} {\bibinfo {author} {\bibfnamefont {A.}~\bibnamefont
  {Bermudez}}, \bibinfo {author} {\bibfnamefont {P.}~\bibnamefont {Schindler}},
  \bibinfo {author} {\bibfnamefont {T.}~\bibnamefont {Monz}}, \bibinfo {author}
  {\bibfnamefont {R.}~\bibnamefont {Blatt}}, \ and\ \bibinfo {author}
  {\bibfnamefont {M.}~\bibnamefont {Müller}},\ }\href {\doibase
  10.1088/1367-2630/aa86eb} {\bibfield  {journal} {\bibinfo  {journal} {New
  Journal of Physics}\ }\textbf {\bibinfo {volume} {19}},\ \bibinfo {pages}
  {113038} (\bibinfo {year} {2017})}\BibitemShut {NoStop}%
\bibitem [{\citenamefont {Ratcliffe}\ \emph {et~al.}(2020)\citenamefont
  {Ratcliffe}, \citenamefont {Oberg},\ and\ \citenamefont
  {Hope}}]{PhysRevA.101.052332}%
  \BibitemOpen
  \bibfield  {author} {\bibinfo {author} {\bibfnamefont {A.~K.}\ \bibnamefont
  {Ratcliffe}}, \bibinfo {author} {\bibfnamefont {L.~M.}\ \bibnamefont
  {Oberg}}, \ and\ \bibinfo {author} {\bibfnamefont {J.~J.}\ \bibnamefont
  {Hope}},\ }\href {\doibase 10.1103/PhysRevA.101.052332} {\bibfield  {journal}
  {\bibinfo  {journal} {Phys. Rev. A}\ }\textbf {\bibinfo {volume} {101}},\
  \bibinfo {pages} {052332} (\bibinfo {year} {2020})}\BibitemShut {NoStop}%
\bibitem [{\citenamefont {Wu}\ \emph {et~al.}(2021{\natexlab{b}})\citenamefont
  {Wu}, \citenamefont {Liu}, \citenamefont {Zhao},\ and\ \citenamefont
  {Duan}}]{PhysRevA.103.022419}%
  \BibitemOpen
  \bibfield  {author} {\bibinfo {author} {\bibfnamefont {Y.-K.}\ \bibnamefont
  {Wu}}, \bibinfo {author} {\bibfnamefont {Z.-D.}\ \bibnamefont {Liu}},
  \bibinfo {author} {\bibfnamefont {W.-D.}\ \bibnamefont {Zhao}}, \ and\
  \bibinfo {author} {\bibfnamefont {L.-M.}\ \bibnamefont {Duan}},\ }\href
  {\doibase 10.1103/PhysRevA.103.022419} {\bibfield  {journal} {\bibinfo
  {journal} {Phys. Rev. A}\ }\textbf {\bibinfo {volume} {103}},\ \bibinfo
  {pages} {022419} (\bibinfo {year} {2021}{\natexlab{b}})}\BibitemShut
  {NoStop}%
\bibitem [{\citenamefont {Duan}\ and\ \citenamefont
  {Yang}(2021)}]{patent-ZL202110047218.2}%
  \BibitemOpen
  \bibfield  {author} {\bibinfo {author} {\bibfnamefont {L.-M.}\ \bibnamefont
  {Duan}}\ and\ \bibinfo {author} {\bibfnamefont {H.-X.}\ \bibnamefont
  {Yang}},\ }\href {https://patents.google.com/patent/CN112749808B/en}
  {\enquote {\bibinfo {title} {Addressing control system and addressing control
  method},}\ } (\bibinfo {year} {2021})\BibitemShut {NoStop}%
\bibitem [{\citenamefont {Pu}\ \emph {et~al.}(2017)\citenamefont {Pu},
  \citenamefont {Jiang}, \citenamefont {Chang}, \citenamefont {Yang},
  \citenamefont {Li},\ and\ \citenamefont {Duan}}]{pu2017experimental}%
  \BibitemOpen
  \bibfield  {author} {\bibinfo {author} {\bibfnamefont {Y.}~\bibnamefont
  {Pu}}, \bibinfo {author} {\bibfnamefont {N.}~\bibnamefont {Jiang}}, \bibinfo
  {author} {\bibfnamefont {W.}~\bibnamefont {Chang}}, \bibinfo {author}
  {\bibfnamefont {H.}~\bibnamefont {Yang}}, \bibinfo {author} {\bibfnamefont
  {C.}~\bibnamefont {Li}}, \ and\ \bibinfo {author} {\bibfnamefont
  {L.}~\bibnamefont {Duan}},\ }\href {https://doi.org/10.1038/ncomms15359}
  {\bibfield  {journal} {\bibinfo  {journal} {Nature communications}\ }\textbf
  {\bibinfo {volume} {8}},\ \bibinfo {pages} {15359} (\bibinfo {year}
  {2017})}\BibitemShut {NoStop}%
\bibitem [{\citenamefont {Wang}\ \emph
  {et~al.}(2020{\natexlab{b}})\citenamefont {Wang}, \citenamefont {Crain},
  \citenamefont {Fang}, \citenamefont {Zhang}, \citenamefont {Huang},
  \citenamefont {Liang}, \citenamefont {Leung}, \citenamefont {Brown},\ and\
  \citenamefont {Kim}}]{PhysRevLett.125.150505}%
  \BibitemOpen
  \bibfield  {author} {\bibinfo {author} {\bibfnamefont {Y.}~\bibnamefont
  {Wang}}, \bibinfo {author} {\bibfnamefont {S.}~\bibnamefont {Crain}},
  \bibinfo {author} {\bibfnamefont {C.}~\bibnamefont {Fang}}, \bibinfo {author}
  {\bibfnamefont {B.}~\bibnamefont {Zhang}}, \bibinfo {author} {\bibfnamefont
  {S.}~\bibnamefont {Huang}}, \bibinfo {author} {\bibfnamefont
  {Q.}~\bibnamefont {Liang}}, \bibinfo {author} {\bibfnamefont {P.~H.}\
  \bibnamefont {Leung}}, \bibinfo {author} {\bibfnamefont {K.~R.}\ \bibnamefont
  {Brown}}, \ and\ \bibinfo {author} {\bibfnamefont {J.}~\bibnamefont {Kim}},\
  }\href {\doibase 10.1103/PhysRevLett.125.150505} {\bibfield  {journal}
  {\bibinfo  {journal} {Phys. Rev. Lett.}\ }\textbf {\bibinfo {volume} {125}},\
  \bibinfo {pages} {150505} (\bibinfo {year} {2020}{\natexlab{b}})}\BibitemShut
  {NoStop}%
\bibitem [{\citenamefont {Barredo}\ \emph {et~al.}(2018)\citenamefont
  {Barredo}, \citenamefont {Lienhard}, \citenamefont {de~L{\'e}s{\'e}leuc},
  \citenamefont {Lahaye},\ and\ \citenamefont {Browaeys}}]{Barredo2018}%
  \BibitemOpen
  \bibfield  {author} {\bibinfo {author} {\bibfnamefont {D.}~\bibnamefont
  {Barredo}}, \bibinfo {author} {\bibfnamefont {V.}~\bibnamefont {Lienhard}},
  \bibinfo {author} {\bibfnamefont {S.}~\bibnamefont {de~L{\'e}s{\'e}leuc}},
  \bibinfo {author} {\bibfnamefont {T.}~\bibnamefont {Lahaye}}, \ and\ \bibinfo
  {author} {\bibfnamefont {A.}~\bibnamefont {Browaeys}},\ }\href {\doibase
  10.1038/s41586-018-0450-2} {\bibfield  {journal} {\bibinfo  {journal}
  {Nature}\ }\textbf {\bibinfo {volume} {561}},\ \bibinfo {pages} {79}
  (\bibinfo {year} {2018})}\BibitemShut {NoStop}%
\bibitem [{\citenamefont {Ebadi}\ \emph {et~al.}(2021)\citenamefont {Ebadi},
  \citenamefont {Wang}, \citenamefont {Levine}, \citenamefont {Keesling},
  \citenamefont {Semeghini}, \citenamefont {Omran}, \citenamefont {Bluvstein},
  \citenamefont {Samajdar}, \citenamefont {Pichler}, \citenamefont {Ho},
  \citenamefont {Choi}, \citenamefont {Sachdev}, \citenamefont {Greiner},
  \citenamefont {Vuleti{\'{c}}},\ and\ \citenamefont {Lukin}}]{Ebadi2021}%
  \BibitemOpen
  \bibfield  {author} {\bibinfo {author} {\bibfnamefont {S.}~\bibnamefont
  {Ebadi}}, \bibinfo {author} {\bibfnamefont {T.~T.}\ \bibnamefont {Wang}},
  \bibinfo {author} {\bibfnamefont {H.}~\bibnamefont {Levine}}, \bibinfo
  {author} {\bibfnamefont {A.}~\bibnamefont {Keesling}}, \bibinfo {author}
  {\bibfnamefont {G.}~\bibnamefont {Semeghini}}, \bibinfo {author}
  {\bibfnamefont {A.}~\bibnamefont {Omran}}, \bibinfo {author} {\bibfnamefont
  {D.}~\bibnamefont {Bluvstein}}, \bibinfo {author} {\bibfnamefont
  {R.}~\bibnamefont {Samajdar}}, \bibinfo {author} {\bibfnamefont
  {H.}~\bibnamefont {Pichler}}, \bibinfo {author} {\bibfnamefont {W.~W.}\
  \bibnamefont {Ho}}, \bibinfo {author} {\bibfnamefont {S.}~\bibnamefont
  {Choi}}, \bibinfo {author} {\bibfnamefont {S.}~\bibnamefont {Sachdev}},
  \bibinfo {author} {\bibfnamefont {M.}~\bibnamefont {Greiner}}, \bibinfo
  {author} {\bibfnamefont {V.}~\bibnamefont {Vuleti{\'{c}}}}, \ and\ \bibinfo
  {author} {\bibfnamefont {M.~D.}\ \bibnamefont {Lukin}},\ }\href {\doibase
  10.1038/s41586-021-03582-4} {\bibfield  {journal} {\bibinfo  {journal}
  {Nature}\ }\textbf {\bibinfo {volume} {595}},\ \bibinfo {pages} {227}
  (\bibinfo {year} {2021})}\BibitemShut {NoStop}%
\bibitem [{\citenamefont {Roman}\ \emph {et~al.}(2020)\citenamefont {Roman},
  \citenamefont {Ransford}, \citenamefont {Ip},\ and\ \citenamefont
  {Campbell}}]{Roman2020}%
  \BibitemOpen
  \bibfield  {author} {\bibinfo {author} {\bibfnamefont {C.}~\bibnamefont
  {Roman}}, \bibinfo {author} {\bibfnamefont {A.}~\bibnamefont {Ransford}},
  \bibinfo {author} {\bibfnamefont {M.}~\bibnamefont {Ip}}, \ and\ \bibinfo
  {author} {\bibfnamefont {W.~C.}\ \bibnamefont {Campbell}},\ }\href {\doibase
  10.1088/1367-2630/ab9982} {\bibfield  {journal} {\bibinfo  {journal} {New
  Journal of Physics}\ }\textbf {\bibinfo {volume} {22}},\ \bibinfo {pages}
  {073038} (\bibinfo {year} {2020})}\BibitemShut {NoStop}%
\bibitem [{\citenamefont {Edmunds}\ \emph {et~al.}(2021)\citenamefont
  {Edmunds}, \citenamefont {Tan}, \citenamefont {Milne}, \citenamefont {Singh},
  \citenamefont {Biercuk},\ and\ \citenamefont {Hempel}}]{edmunds2020scalable}%
  \BibitemOpen
  \bibfield  {author} {\bibinfo {author} {\bibfnamefont {C.~L.}\ \bibnamefont
  {Edmunds}}, \bibinfo {author} {\bibfnamefont {T.~R.}\ \bibnamefont {Tan}},
  \bibinfo {author} {\bibfnamefont {A.~R.}\ \bibnamefont {Milne}}, \bibinfo
  {author} {\bibfnamefont {A.}~\bibnamefont {Singh}}, \bibinfo {author}
  {\bibfnamefont {M.~J.}\ \bibnamefont {Biercuk}}, \ and\ \bibinfo {author}
  {\bibfnamefont {C.}~\bibnamefont {Hempel}},\ }\href {\doibase
  10.1103/PhysRevA.104.012606} {\bibfield  {journal} {\bibinfo  {journal}
  {Phys. Rev. A}\ }\textbf {\bibinfo {volume} {104}},\ \bibinfo {pages}
  {012606} (\bibinfo {year} {2021})}\BibitemShut {NoStop}%
\bibitem [{\citenamefont {Green}\ and\ \citenamefont
  {Biercuk}(2015)}]{greenPhaseModulatedDecouplingError2015}%
  \BibitemOpen
  \bibfield  {author} {\bibinfo {author} {\bibfnamefont {T.~J.}\ \bibnamefont
  {Green}}\ and\ \bibinfo {author} {\bibfnamefont {M.~J.}\ \bibnamefont
  {Biercuk}},\ }\href {\doibase 10.1103/PhysRevLett.114.120502} {\bibfield
  {journal} {\bibinfo  {journal} {Physical Review Letters}\ }\textbf {\bibinfo
  {volume} {114}},\ \bibinfo {pages} {120502} (\bibinfo {year}
  {2015})}\BibitemShut {NoStop}%
\bibitem [{\citenamefont {Debnath}(2016)}]{debnath2016programmable}%
  \BibitemOpen
  \bibfield  {author} {\bibinfo {author} {\bibfnamefont {S.}~\bibnamefont
  {Debnath}},\ }\emph {\bibinfo {title} {A programmable five qubit quantum
  computer using trapped atomic ions}},\ \href@noop {} {Ph.D. thesis},\
  \bibinfo  {school} {University of Maryland, College Park} (\bibinfo {year}
  {2016})\BibitemShut {NoStop}%
\bibitem [{\citenamefont {Jia}\ \emph {et~al.}(2023)\citenamefont {Jia},
  \citenamefont {Huang}, \citenamefont {Kang}, \citenamefont {Sun},
  \citenamefont {Spivey}, \citenamefont {Kim},\ and\ \citenamefont
  {Brown}}]{PhysRevA.107.032617}%
  \BibitemOpen
  \bibfield  {author} {\bibinfo {author} {\bibfnamefont {Z.}~\bibnamefont
  {Jia}}, \bibinfo {author} {\bibfnamefont {S.}~\bibnamefont {Huang}}, \bibinfo
  {author} {\bibfnamefont {M.}~\bibnamefont {Kang}}, \bibinfo {author}
  {\bibfnamefont {K.}~\bibnamefont {Sun}}, \bibinfo {author} {\bibfnamefont
  {R.~F.}\ \bibnamefont {Spivey}}, \bibinfo {author} {\bibfnamefont
  {J.}~\bibnamefont {Kim}}, \ and\ \bibinfo {author} {\bibfnamefont {K.~R.}\
  \bibnamefont {Brown}},\ }\href {\doibase 10.1103/PhysRevA.107.032617}
  {\bibfield  {journal} {\bibinfo  {journal} {Phys. Rev. A}\ }\textbf {\bibinfo
  {volume} {107}},\ \bibinfo {pages} {032617} (\bibinfo {year}
  {2023})}\BibitemShut {NoStop}%
\bibitem [{\citenamefont {{Shi-Liang Zhu}}\ \emph {et~al.}(2006)\citenamefont
  {{Shi-Liang Zhu}}, \citenamefont {{C. Monroe}},\ and\ \citenamefont {{L.-M.
  Duan}}}]{Zhu2006}%
  \BibitemOpen
  \bibfield  {author} {\bibinfo {author} {\bibnamefont {{Shi-Liang Zhu}}},
  \bibinfo {author} {\bibnamefont {{C. Monroe}}}, \ and\ \bibinfo {author}
  {\bibnamefont {{L.-M. Duan}}},\ }\href {\doibase 10.1209/epl/i2005-10424-4}
  {\bibfield  {journal} {\bibinfo  {journal} {Europhys. Lett.}\ }\textbf
  {\bibinfo {volume} {73}},\ \bibinfo {pages} {485} (\bibinfo {year}
  {2006})}\BibitemShut {NoStop}%
\bibitem [{\citenamefont {Leung}\ and\ \citenamefont
  {Brown}(2018)}]{PhysRevA.98.032318}%
  \BibitemOpen
  \bibfield  {author} {\bibinfo {author} {\bibfnamefont {P.~H.}\ \bibnamefont
  {Leung}}\ and\ \bibinfo {author} {\bibfnamefont {K.~R.}\ \bibnamefont
  {Brown}},\ }\href {\doibase 10.1103/PhysRevA.98.032318} {\bibfield  {journal}
  {\bibinfo  {journal} {Phys. Rev. A}\ }\textbf {\bibinfo {volume} {98}},\
  \bibinfo {pages} {032318} (\bibinfo {year} {2018})}\BibitemShut {NoStop}%
\bibitem [{\citenamefont {Leibfried}\ \emph {et~al.}(2003)\citenamefont
  {Leibfried}, \citenamefont {Blatt}, \citenamefont {Monroe},\ and\
  \citenamefont {Wineland}}]{leibfriedQuantumDynamicsSingle2003}%
  \BibitemOpen
  \bibfield  {author} {\bibinfo {author} {\bibfnamefont {D.}~\bibnamefont
  {Leibfried}}, \bibinfo {author} {\bibfnamefont {R.}~\bibnamefont {Blatt}},
  \bibinfo {author} {\bibfnamefont {C.}~\bibnamefont {Monroe}}, \ and\ \bibinfo
  {author} {\bibfnamefont {D.}~\bibnamefont {Wineland}},\ }\href {\doibase
  10.1103/RevModPhys.75.281} {\bibfield  {journal} {\bibinfo  {journal}
  {Reviews of Modern Physics}\ }\textbf {\bibinfo {volume} {75}},\ \bibinfo
  {pages} {281} (\bibinfo {year} {2003})}\BibitemShut {NoStop}%
\end{thebibliography}
%

\end{document}


\title{Supplementary Information for ``Individually Addressed Entangling Gates in a Two-Dimensional Ion Crystal''}

\author{Y.-H. Hou}
\thanks{These authors contribute equally to this work}%
\affiliation{Center for Quantum Information, Institute for Interdisciplinary Information Sciences, Tsinghua University, Beijing 100084, PR China}

\author{Y.-J. Yi}
\thanks{These authors contribute equally to this work}%
\affiliation{Center for Quantum Information, Institute for Interdisciplinary Information Sciences, Tsinghua University, Beijing 100084, PR China}

\author{Y.-K. Wu}
\thanks{These authors contribute equally to this work}%
\affiliation{Center for Quantum Information, Institute for Interdisciplinary Information Sciences, Tsinghua University, Beijing 100084, PR China}
\affiliation{Hefei National Laboratory, Hefei 230088, PR China}

\author{Y.-Y. Chen}
\affiliation{Center for Quantum Information, Institute for Interdisciplinary Information Sciences, Tsinghua University, Beijing 100084, PR China}

\author{L. Zhang}
\affiliation{Center for Quantum Information, Institute for Interdisciplinary Information Sciences, Tsinghua University, Beijing 100084, PR China}

\author{Y. Wang}
\affiliation{Center for Quantum Information, Institute for Interdisciplinary Information Sciences, Tsinghua University, Beijing 100084, PR China}
\affiliation{HYQ Co., Ltd., Beijing 100176, PR China}

\author{Y.-L. Xu}
\affiliation{Center for Quantum Information, Institute for Interdisciplinary Information Sciences, Tsinghua University, Beijing 100084, PR China}

\author{C. Zhang}
\affiliation{Center for Quantum Information, Institute for Interdisciplinary Information Sciences, Tsinghua University, Beijing 100084, PR China}
\affiliation{HYQ Co., Ltd., Beijing 100176, PR China}

\author{Q.-X. Mei}
\affiliation{HYQ Co., Ltd., Beijing 100176, PR China}

\author{H.-X. Yang}
\affiliation{HYQ Co., Ltd., Beijing 100176, PR China}

\author{J.-Y. Ma}
\affiliation{HYQ Co., Ltd., Beijing 100176, PR China}

\author{S.-A. Guo}
\affiliation{Center for Quantum Information, Institute for Interdisciplinary Information Sciences, Tsinghua University, Beijing 100084, PR China}

\author{J. Ye}
\affiliation{Center for Quantum Information, Institute for Interdisciplinary Information Sciences, Tsinghua University, Beijing 100084, PR China}

\author{B.-X. Qi}
\affiliation{Center for Quantum Information, Institute for Interdisciplinary Information Sciences, Tsinghua University, Beijing 100084, PR China}

\author{Z.-C. Zhou}
\affiliation{Center for Quantum Information, Institute for Interdisciplinary Information Sciences, Tsinghua University, Beijing 100084, PR China}
\affiliation{Hefei National Laboratory, Hefei 230088, PR China}

\author{P.-Y. Hou}
\affiliation{Center for Quantum Information, Institute for Interdisciplinary Information Sciences, Tsinghua University, Beijing 100084, PR China}
\affiliation{Hefei National Laboratory, Hefei 230088, PR China}

\author{L.-M. Duan}
\email{lmduan@tsinghua.edu.cn}
\affiliation{Center for Quantum Information, Institute for Interdisciplinary Information Sciences, Tsinghua University, Beijing 100084, PR China}
\affiliation{Hefei National Laboratory, Hefei 230088, PR China}
\affiliation{New Cornerstone Science Laboratory, Beijing 100084, PR China}

\maketitle

\section{Blade trap for 2D ion crystals with large optical access}

\begin{figure*}[htbp]
	\centering
	\includegraphics[width=0.5\linewidth]{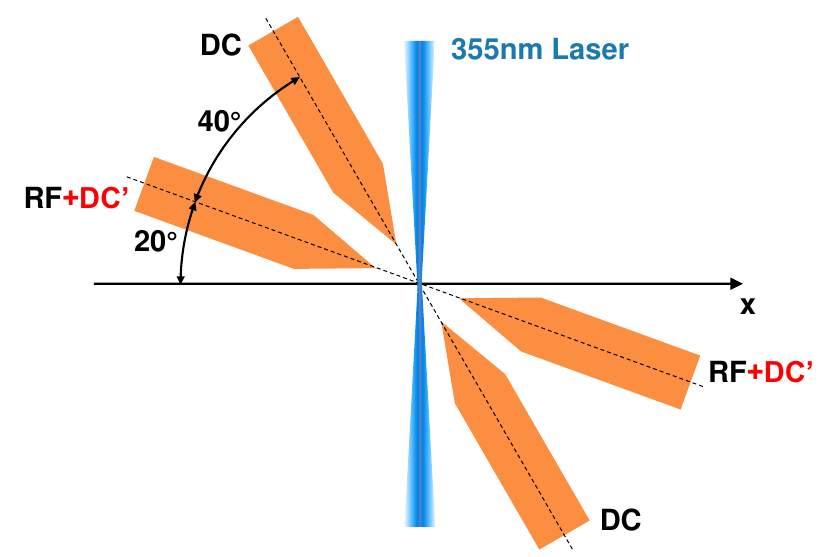}
	\caption {Side view of our blade trap for 2D ion crystals. A DC bias (DC$^\prime$) is applied on the RF electrodes to split the radial trap frequencies. The 2D ion crystal locates close to the $xz$ plane, perpendicular to the counter-propagating $355\,$nm Raman laser beams and the imaging system in the $y$ direction.}
	\label{figs_trap}
\end{figure*}

Our blade trap for 2D ion crystals is sketched in Fig.~\ref{figs_trap}. The DC and the RF blades are at an angle of $40^\circ$, with their inner edges separated along the diagonal direction by $500\,\mu$m.
With this configuration, when we apply zero voltage on the DC electrodes (thus no axial trapping), the two radial principal axes of the trap will be $45^\circ$ from the symmetric axes of these blades, or about $5^\circ$ from the $x$ and $y$ axes in the plot. We further apply a DC bias on the RF electrodes to split the two radial trap frequencies such that $\omega_y\gg\omega_x$. Finally we introduce a weaker axial trapping $\omega_z$ by the voltage pattern on the five segments of each DC blade, which does not significantly change the radial principal axes. Therefore the obtained 2D crystal will locate close to the $xz$ plane.
This design gives us larger optical access to the ion crystal than the typical monolithic 3D ion traps. Apart from the counter-propagating Raman laser beams and the imaging system perpendicular to the ion plane, our system also allows a detection laser along the micromotion-free $z$ direction and a Doppler cooling laser at $20^\circ$ from the $xz$ plane with equal angles to the $x$ and $z$ axes.

\section{Correction of detection error}
We use an EMCCD camera to collect the fluorescence from individual ions under $370\,$nm detection laser for a $1\,$ms duration. Due to the low NA of $0.34$ of our homemade objective, the quantum efficiency of the EMCCD, and the loss on many optic elements, we have a relatively low photon collection efficiency, which restricts the single-shot detection fidelity. From the measured photon count distributions for the bright and the dark states, we estimate a detection infidelity of about $7\%$ for each ion.
Moreover, aberrations in the imaging system increase the size of each ion's image and cause their overlap. This leads to crosstalk between different ions. In other words, an ion is more likely to be detected as bright when its adjacent ions are in the bright state. To calibrate this error, we prepare all the ions in $\ket{+}$ by optical pumping and a microwave $\pi/2$ pulse. Ideally there should be no correlation in the measured states for different ions, while experimentally we estimate an average detection crosstalk of about $1\%$ from the measured correlations.

Although such errors limit the performance of single-shot measurements, we can still recover the probability distribution in the computational basis when only the expectation values are needed. For this purpose, first we prepare all the $16$ computational basis states for the four-ion crystal by optical pumping into $\ket{0000}$ followed by a global microwave $\pi$ pulse and/or several individually addressed single-qubit $\pi$ pulses. To suppress the slow drift in the laser intensity and to enhance the single-qubit gate fidelity, we use the Tycko's composite pulse sequence which consists of three sequential $\pi$ pulses \cite{PhysRevLett.51.775}.
For each computational basis state, we repeat the preparation-detection sequence for $10000$ times to estimate its distribution under the detection errors. This gives us a $16\times 16$ matrix $M$ whose columns are the distributions for each computational basis state.

Now for any quantum state to be measured, we obtain a vector of frequencies $\boldsymbol{f}=[f_0,f_1,\cdots,f_{15}]^T$ over the $16$ possible states from the total $T=\sum_i f_i$ repetitions. Our task is to find the most likely probability distribution $\boldsymbol{p}=[p_0,p_1,\cdots,p_{15}]^T$ such that the distribution $M\boldsymbol{p}$ can generate the observed frequencies $\boldsymbol{f}$ following a multinomial distribution. This maximum likelihood estimation problem can be solved by numerically optimizing $\boldsymbol{p}$ under the constraint that all the probabilities are nonnegative. We further estimate the error bar of the recovered observable by Monte Carlo sampling from the theoretical probability distribution of $\boldsymbol{f}/T$ and computing the standard deviation of the simulated observables. For the data presented in the main text, we measure the population like Fig.~2(a) for $T=2000$ repetitions, and we measure each data point for the Rabi oscillation like Fig.~1(c) and the parity oscillation like Fig.~2(b) for $T=200$ repetitions.

\section{Phase-modulated gate design}
Our two-qubit entangling gates are realized by counter-propagating bichromatic $355\,$nm Raman laser beams with symmetric blue- and red-detuned frequency components \cite{debnath2016programmable}. We have a Hamiltonian in the form of a spin-dependent force as $H=\sum_{jk} (\eta_k b_{jk}\Omega_j/2)\{a_k e^{i[(\mu-\omega_k)t+\phi_m^j]} + h.c.\}\sigma_{\phi_s^j}^j$ where $j$ goes over all the target ions, $k$ goes over all the phonon modes, $\eta_k$ is the Lamb-Dicke parameter of Mode $k$ with a frequency $\omega_k$ and an annihilation operator $a_k$, $b_{jk}$ is the mode vector of Ion $j$ in Mode $k$, $\Omega_j$ is the Raman Rabi frequency on Ion $j$, and $\mu$ is the symmetric detuning of the Raman laser. The motional phase $\phi_m^j=(\varphi_b^j-\varphi_r^j)/2$ and the spin phase $\phi_s^j=(\varphi_b^j+\varphi_r^j)/2$ are determined by the phases of the blue- and red-detuned components $\varphi_b^j$ and $\varphi_r^j$ on Ion $j$, and the spin operator on Ion $j$ is defined as $\sigma_{\phi_s^j}^j=\sigma_x^i\cos\phi_s^j + \sigma_y^i\sin\phi_s^j$. By tuning $\varphi_b^j$ and $\varphi_r^j$ simultaneously, we can adjust the motional phase $\phi_m^j$ while keeping the spin phase $\phi_s^j$ a constant, which we define as the $\sigma_x$ axis of the qubit. In the above Hamiltonian we have also performed rotating-wave approximation to drop the far-off-resonant terms.

The unitary evolution under the above Hamiltonian can be expressed as spin-dependent displacements of all the phonon modes together with two-qubit phases between the target ions $U=\prod_{i<j} e^{i\Theta_{ij}\sigma_x^i\sigma_x^j}\prod_k D_k(\sum_j\alpha_{jk}\sigma_x^j)$. The displacement due to Ion $j$ on Mode $k$ is given by
\begin{equation}
\alpha_{jk} = \frac{1}{2i} \eta_k b_{jk} \int \Omega_j e^{-i[(\mu-\omega_k)t+\phi_m^j]} dt,
\end{equation}
which we want to suppress at the end of the gate. The two-qubit phase between ions $i$ and $j$ is given by
\begin{equation}
\Theta_{ij} = -\sum_k \mathrm{Im}\left[\int \alpha_{ik}(t) d\alpha_{jk}^*(t) + \int \alpha_{jk}(t) d\alpha_{ik}^*(t)\right], \label{eq:phase}
\end{equation}
which appears from applying a new displacement $d\alpha$ to the accumulative one $\alpha(t)$ at different directions.

In this experiment we use a phase-modulated gate sequence where we divide the whole gate into several equal segments and set the motional phase $\phi_m^j$ to be piecewise-constant on these segments. When we set the duration of each segment to be a multiple of $2\pi/|\mu-\omega_k|$ for some mode $k$, clearly the spin-dependent displacement vanishes independent of $\phi_m^j$. For the LR and the UD ion pairs, we further cancel the spin-dependent displacement of the COM mode by a phase shift of $\Delta\phi=\pi-(\mu-\omega_1)\Delta T$ between the two segments where $\Delta T$ is the time difference between the starting points of the two segments, namely the duration of a single segment.

As for the LU ion pair, after designing each segment to disentangle Mode 2 and Mode 4 as described in the main text, we still need to choose the phases of different segments to set the spin-dependent displacements of Mode 1 and Mode 3 to zero. This can be achieved using $2^2=4$ segments \cite{greenPhaseModulatedDecouplingError2015} in the pattern of $[0, \pi - (\mu-\omega_3)\times\Delta T, \pi - (\mu-\omega_1)\times 2\Delta T, - (\mu-\omega_3)\times\Delta T- (\mu-\omega_1)\times 2\Delta T]$. Note that here we use the same phase sequence alternatingly on the two target ions, so we have $\Delta T=2[4\pi/(\omega_2-\omega_4)+\Delta t]$ covering two segments and separations. Such a phase sequence can accumulate a nonzero two-qubit phase between the two target ions according to Eq.~(\ref{eq:phase}), and we can scale the overall Raman Rabi rate of the laser to set this phase to $\pm\pi/4$, which gives us the desired maximal entangling gate.

Finally, note that when applying the $355\,$nm Raman laser, there will be a differential AC Stark shift on the target ions on the order of kHz \cite{debnath2016programmable}. On the one hand, we can shift the blue- and red-detuned frequency components accordingly so that we still get a spin-dependent force in a suitable rotating frame. On the other hand, when switching the addressing beam between the two target ions, we are effectively switching between two frames and we need a shift in the spin phase $\phi_s^j$ to compensate it.

\section{Gate error sources}
To estimate the influence of various error sources, we calibrate their strength by separated single-ion or multi-ion experiments. The laser dephasing time is measured to be $\tau_s=4\,$ms by fitting the exponential decay of the Ramsey fringes under the counter-propagating Raman $\pi/2$ pulses. To separate its effect from the measurement of the motional dephasing time, we use the following sequence: First we initialize the ion in $\ket{0}\ket{0}$ of the spin and the motional states by sideband cooling and optical pumping; Then we perform a carrier Raman $\pi/2$ pulse followed by a red-sideband Raman $\pi$ pulse to prepare the superposition of the motional state $\frac{1}{\sqrt{2}}\ket{0}(\ket{0}+\ket{1})$; After waiting for time $t$ to accumulate some unknown phase between the motional states $\ket{0}$ and $\ket{1}$ due to the fluctuation of the trap frequency, we apply another red-sideband $\pi$ pulse and another carrier $\pi/2$ pulse to turn this phase information into the population of the spin. Note that in this way the optical phase is cancelled between the adjacent carrier and red-sideband pulses, and we fit a motional dephasing time of $\tau_m=3\,$ms.

We estimate the average phonon number of the four normal modes in the $y$ direction by comparing the excitation probability of the red and the blue motional sidebands \cite{leibfriedQuantumDynamicsSingle2003} under a weak individually addressed Raman laser. Then by fitting the increase of the average phonon number vs. the waiting time, we obtain the heating rate of the COM mode to be $120\,\mathrm{s}^{-1}$, and those of the other modes to be below $10\,\mathrm{s}^{-1}$.

The above error sources are modeled as Lindblad operators in our numerical analysis and are solved by QuTip \cite{JOHANSSON20131234}. As for the fluctuation of the laser intensity, we treat it as a low-frequency shot-to-shot variation. To estimate its strength, we set the Raman laser to a large detuning $\Delta$ such that its effect is mainly an AC Stark shift on the qubit. Then we perform the Ramsey spectroscopy using microwave $\pi/2$ pulses with this far-detuned Raman laser turned on during the waiting time. The intensity fluctuation thus translates into the phase noise and can be fitted from the Gaussian envelope of the Ramsey fringes. By comparing this decay rate with the oscillation frequency of the Ramsey fringes which is proportional to $\Omega^2/\Delta$, we estimate a standard deviation of $\sigma=1\%$ in the relative Raman Rabi frequency. Its effect on the gate fidelity will further be squared as $(\pi^2/4)\sigma^2$.

\section{Alternating two-qubit gates in large ion crystals}
In the main text, we use phase-modulated pulse sequences to disentangle the spin and the phonon modes exactly. However, its required segment number increases exponentially with the number of phonon modes to be decoupled \cite{greenPhaseModulatedDecouplingError2015}, which is inefficient for large ion crystals. Fortunately, previous works already show that we can use a much smaller number of segments to disentangle the spin and the phonon modes approximately, while still achieving high gate fidelities (see, e.g. Refs.~\cite{Zhu2006,PhysRevA.100.022332}). In principle we can use any degrees of freedom like the amplitude, phase or frequency of the laser to optimize the gate performance. In practice, finding an amplitude-modulated gate sequence is often easier: The direct optimization of gate infidelity can be formulated as a generalized eigenvalue problem, and even with the robustness criteria included, it can still be expressed as an optimization of polynomial functions \cite{PhysRevA.103.022419}. Therefore, here we present results for an alternating gate sequence on two target ions with amplitude modulation.

Following the derivation of Ref.~\cite{PhysRevA.103.022419}, we divide the total gate time into $n_{\mathrm{seg}}$ equal segments with piecewise-constant Raman Rabi rate given by a real vector $\boldsymbol{\Omega} = (\Omega_1, \Omega_2, \cdots, \Omega_{n_{\textrm{seg}}})^T$. The optimization of the spin-dependent displacement [Eq.~(1) of the main text] and the two-qubit phase [Eq.~(2) of the main text], as well as their robustness against small drifts in the trap frequency, can then be expressed as a quartic function of $\boldsymbol{\Omega}$ (see Appendix~B of Ref.~\cite{PhysRevA.103.022419})
\begin{equation}
\Omega^T \boldsymbol{M} \Omega + (\Omega^T \boldsymbol{\gamma} \Omega)^2,
\end{equation}
where the matrix elements of $\boldsymbol{M}$ and $\boldsymbol{\gamma}$ come from the time integral in Eqs.~(1) and (2) of the main text on the corresponding segments under unit Rabi rate.

Now for an alternating pulse sequence, we can still write it as a single vector $\boldsymbol{\Omega} = (\Omega_1, \Omega_2, \cdots, \Omega_{n_{\textrm{seg}}})^T$, knowing that this sequence is to be alternatingly applied on the two target ions. To ensure this in the formulation of the optimization problem, we simply set the spin-dependent displacement on the corresponding segments and the two-qubit phase on the corresponding segment pairs to zero, which gives us a modified matrix $\boldsymbol{M}$ and a modified matrix $\boldsymbol{\gamma}$. We optimize this cost function under the constraint that the accumulated two-qubit phase, which is quadratic in $\boldsymbol{\Omega}$, is equal to $\pm \pi/4$.

\begin{figure}[htbp]
	\centering
	\includegraphics[width=\linewidth]{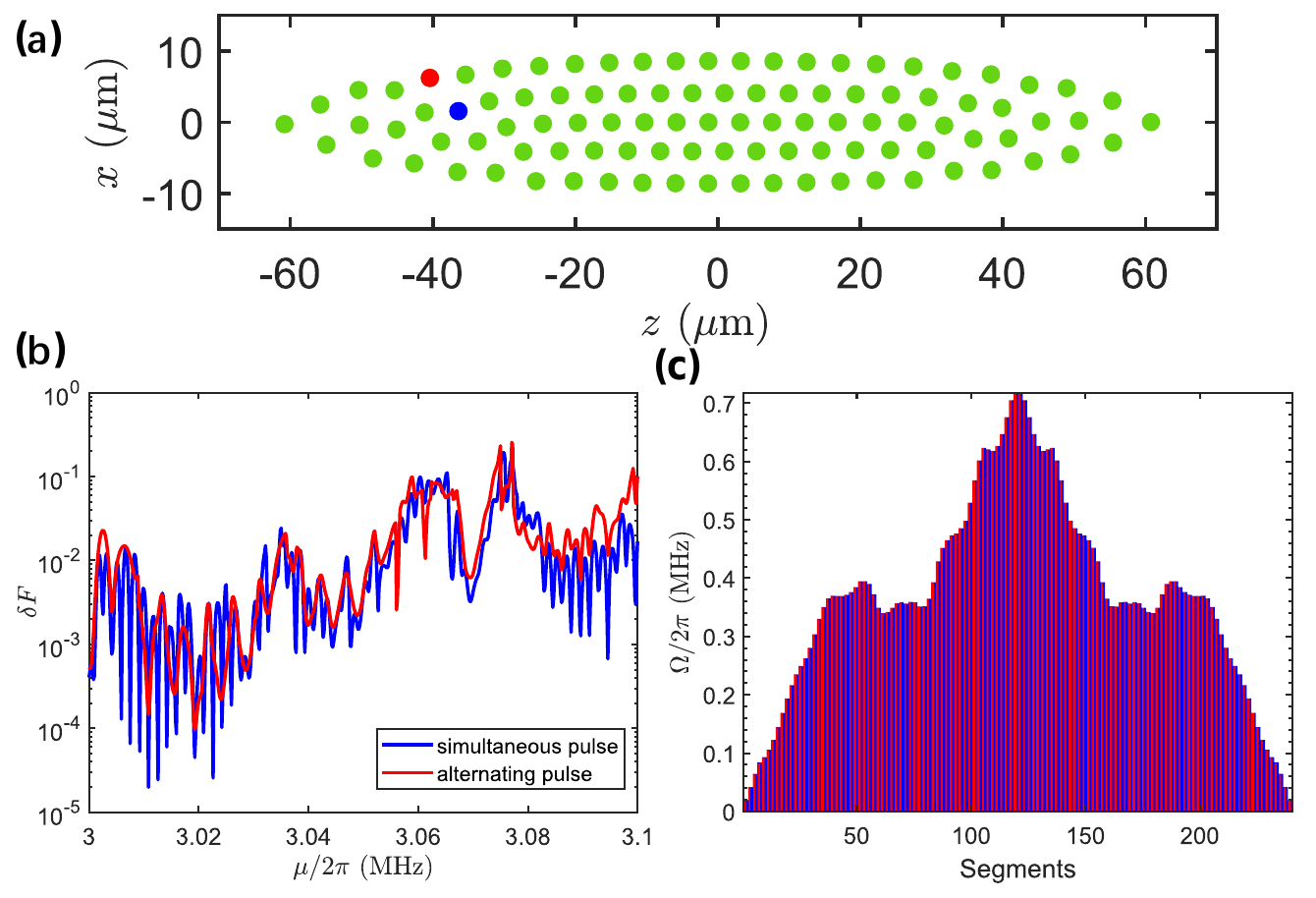}
	\caption {(a) A 2D ion crystal of $N=100$ ions in a harmonic trap with $\omega_x=2\pi\times 0.7\,$MHz, $\omega_y=2\pi\times 3\,$MHz and $\omega_z=2\pi\times 0.2\,$MHz. Two target ions are colored in red and blue, respectively. (b) Optimized gate infidelity for the amplitude-modulated simultaneous-pulse gate (blue) and the alternating-pulse gate (red) vs. the laser detuning $\mu$. We fix a gate time of $300\,\mu$s and use $n_{\mathrm{seg}}=120$ symmetric segments for the simultaneous-pulse gate and $n_{\mathrm{seg}}=240$ symmetric segments for the alternating-pulse gate. (c) Optimized alternating gate sequence at the detuning $\mu=2\pi\times 3.0194\,$MHz, namely $2\pi\times 19.4\,$kHz above the COM mode. The red and blue colors correspond to the laser sequences that are alternatingly applied on the two target ions, respectively. The theoretical gate infidelity is below $10^{-4}$.}
	\label{fig_supp}
\end{figure}

We present some numerical results in Fig.~\ref{fig_supp} for a 2D ion crystal of $N=100$ ions in a harmonic trap with $\omega_x=2\pi\times 0.7\,$MHz, $\omega_y=2\pi\times 3\,$MHz and $\omega_z=2\pi\times 0.2\,$MHz. An actual 2D crystal will be subjected to micromotion, but as shown in Ref.~\cite{guo2023siteresolved}, for even larger crystals, the micromotion amplitude can still be much smaller than the ion spacings such that individual addressing can still be achieved with low crosstalk. Then as we show in the main text, the effect of the micromotion can be compensated by a recalibration of the laser intensity.

We consider two target ions colored in red and blue in Fig.~\ref{fig_supp}(a). We fix a total gate time of $300\,\mu$s and a segment number of $n_{\mathrm{seg}}=240$. Note that here we apply the same amplitude modulation sequence to the two target ions alternatingly and require a time reversal symmetry in the sequence [see Fig.~\ref{fig_supp}(c)]. Therefore, in the $n_{\mathrm{seg}}=240$ segments there are only $n_{\mathrm{seg}}/4=60$ adjustable parameters, smaller than the number of phonon modes to be distentangled exactly. We scan the Raman laser detuning and optimize the gate fidelity as the red curve in Fig.~\ref{fig_supp}(b). We obtain the gate sequence in Fig.~\ref{fig_supp}(c) at the detuning $\mu=2\pi\times 3.0194\,$MHz, namely $2\pi\times 19.4\,$kHz above the COM mode, with a theoretical gate fidelity above $99.99\%$ assuming an average phonon number of $0.5$ for each mode. As a comparison, we also present a gate design when the laser beams are applied on the two target ions simultaneously. For a fair comparison, we use $n_{\mathrm{seg}}=120$ segments so that there are again $60$ adjustable parameters when we require the amplitude modulation sequence to be time-reversible. We set a duration of $1.25\,\mu$s for each segment and wait for another $1.25\,\mu$s between adjacent segments so that the total gate time is still $300\,\mu$s. The result is shown as the blue curve in Fig.~\ref{fig_supp}(b). As we can see, the two curves have similar tendency, which suggests that using alternating gate sequence does not harm the gate performance.

\section{Additional data for crosstalk in individual addressing}
\begin{figure*}[htbp]
	\centering
	\includegraphics[width=0.8\linewidth]{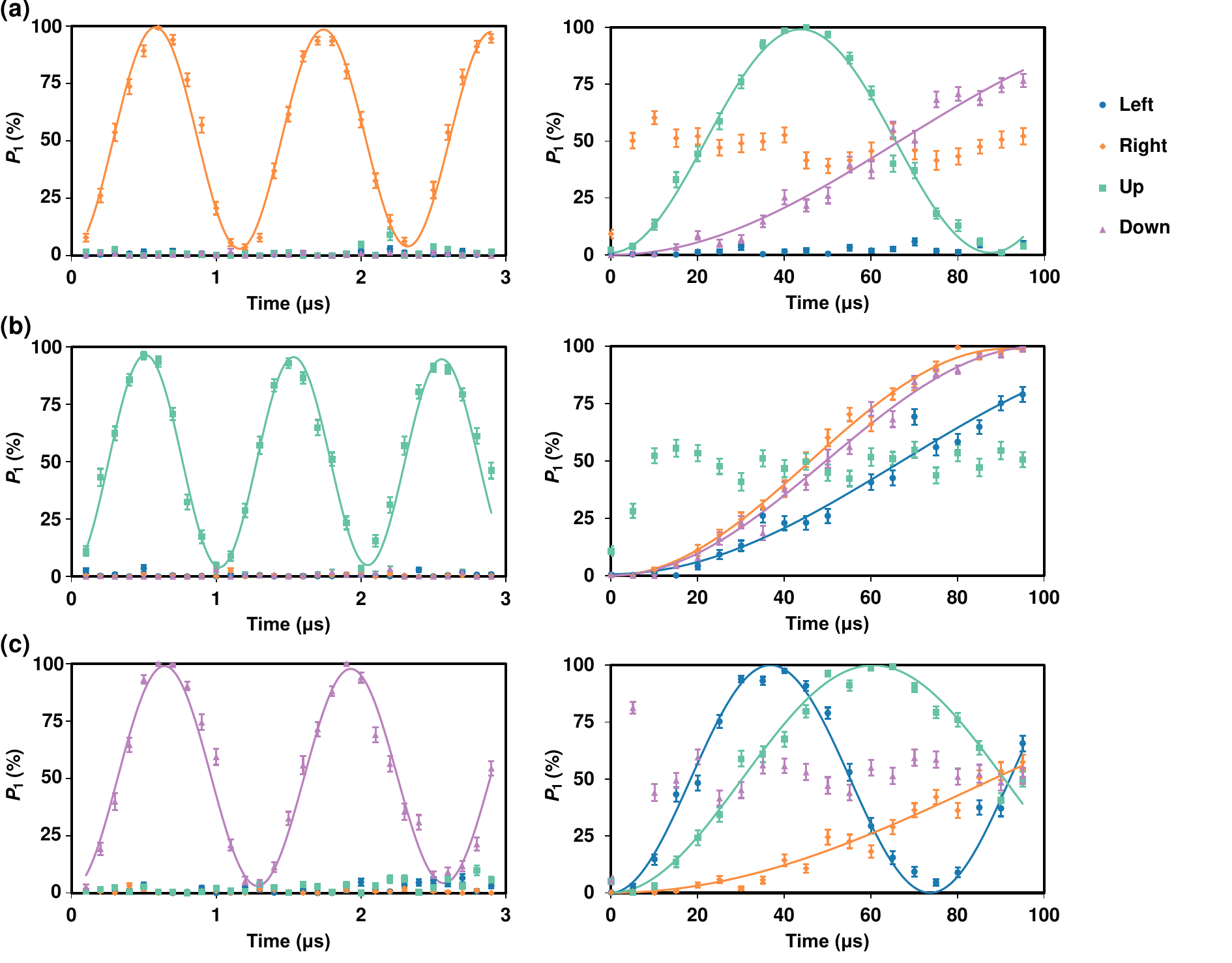}
	\caption {Individual addressing of (a) the right ion, (b) the up ion, and (c) the down ion. The left and the right panels show the short-time and the long-time Rabi oscillations, respectively. Due to the aberration of the addressing beams, the crosstalk errors are not exactly symmetric, but for all the target ions we can bound the crosstalk infidelity for a single-qubit $\pi$ pulse to be below $0.08\%$.}
	\label{figs_addressing}
\end{figure*}

In Fig.~\ref{figs_addressing} we show additional data for individual addressing of different target ions in the four-ion crystal similar to Figs.~1(c) and (d) of the main text.

%